\documentclass[11pt,a4paper]{article}
\pdfoutput=1
\usepackage[utf8x]{inputenc} 
\usepackage{jheppub}
\usepackage{amsfonts}
\usepackage{amsmath}
\usepackage{slashed}
\usepackage{amssymb}
\usepackage{graphicx}
\usepackage{epic}
\usepackage{eepic}
\usepackage{epsfig}
\usepackage{latexsym}
\usepackage{color}
\usepackage{float}
\usepackage{multirow}
\usepackage{hyperref}
\usepackage{enumitem}
\usepackage{tabularx,multirow,array}
\hypersetup{colorlinks=true,citecolor=red,linkcolor=magenta,urlcolor=blue}
\usepackage[caption=false]{subfig}
\usepackage{natbib}
\usepackage{relsize}
\usepackage{breqn}
\usepackage{shorthand}
\usepackage{subfig}
\usepackage[normalem]{ulem}

\def\dis{\displaystyle}
\def\beq{\begin{equation}}
\def\eeq{\end{equation}}
\def\barr{\begin{array}}
\def\earr{\end{array}}

\definecolor{darkred}{cmyk}{0,1,1,0.4}

\usepackage{color}
 \usepackage[normalem]{ulem}
\long\def\/*#1*/{}
\usepackage{color}
 \usepackage[normalem]{ulem}
\definecolor{darkgreen}{cmyk}{1,0,1,0.4}
\definecolor{darkred}{cmyk}{0,1,1,0.4}

\title{Leptogenesis in an anomaly-free $\mathrm{U}(1)$ extension with higher-dimensional operators}

\author[a]{Kuldeep Deka,}
\emailAdd{kuldeepdeka.physics@gmail.com}

\author[b]{Tanumoy Mandal,}
\emailAdd{tanumoy@iisertvm.ac.in}

\author[c]{Ananya Mukherjee,}
\emailAdd{ananyatezpur@gmail.com}

\author[d]{Soumya Sadhukhan}
\emailAdd{physicsoumya@gmail.com}

\affiliation[a]{Department of Physics and Astrophysics, University of Delhi, Delhi 110 007, India}
\affiliation[b]{Indian Institute of Science Education and Research Thiruvananthapuram, Vithura, Kerala, 695551, India}
\affiliation[c]{Department of Physics, University of Calcutta, 92 Acharya Prafulla Chandra Road, Kolkata 700009, India}
\affiliation[d]{Ramakrishna Mission Residential College (Autonomous), Narendrapur, Kolkata 700103, India}

\date{\today}

\abstract{We explore an anomaly-free ${\textrm U}(1)$ gauge extended beyond the Standard model (BSM) framework, to account for the baryon asymmetry of the Universe, along with arranging for tiny neutrino mass. Neutrino masses are generated via higher-dimensional operators (HDOs) involving three right-handed neutrinos (RHNs) with gauge charges ($4$, $4$ and $-5$ respectively) and two BSM scalars. This is an attractive framework as it can accommodate a keV scale dark matter, with the lightest RHN being the candidate. The remaining two RHNs are quasi-degenerate at the TeV-scale, actively participating in the process of resonant leptogenesis through their decay governed by the same set of HDOs. The RHNs being at the TeV scale, make this framework relevant for studying flavored resonant leptogenesis.  This TeV-scale resonant leptogenesis, after satisfying the neutrino oscillation data, leads to interesting predictions on the Yukawa sector of the model. The thermal evolution of the baryon asymmetry has followed the experimental results rather accurately in that corner of parameter space. As a matter of fact, this TeV-scale framework which in principle relies on the low scale resonant leptogenesis typically leads to predictions that potentially can be tested at the colliders.  In particular, we consider the same-sign dilepton signature that arises from the RHN pair production through the decay of heavy gauge boson of the extra ${\textrm U}(1)$.}


\begin{document}

\maketitle
\def\lapp{\mathrel{\rlap{\raise.5ex\hbox{$<$}}
                    {\lower.5ex\hbox{$\sim$}}}}
\def\gapp{\mathrel{\rlap{\raise.5ex\hbox{$>$}}
                    {\lower.5ex\hbox{$\sim$}}}}    
\section{Introduction}
The recent arrival of the evidence of nonzero neutrino masses through the observation of neutrino oscillation \cite{Tanabashi:2018oca,Abe:2018wpn,NOvA:2018gge,Adamson:2013whj,Adey:2018zwh,Bak:2018ydk,Abe:2014bwa,Abe:2017aap} accounts for one of the major findings which calls for beyond the Standard Model~(SM) frameworks. In addition to the origin of neutrino mass, the SM of particle physics also has some other theoretical limitations namely the explanation for the matter-antimatter asymmetry of the Universe (also called the baryon asymmetry of the Universe (BAU)) and the absence of a particle dark matter (DM) candidate. Most interestingly, the neutrino mass generation mechanism can be realized in such a way that all of the aforementioned issues can be well-addressed within a single framework (for some recent works, you may look into~\cite{Borah:2017qdu,Konar:2020vuu,Dey:2021ecr,Klaric:2021cpi,Fong:2021tqj,Domcke:2020quw,Granelli:2020ysj,Klaric:2020lov,Samanta:2020gdw,Das:2020vca,Mahanta:2019sfo,Borah:2018rca,Boruah:2021qlf,Gautam:2020wsd,Jiang:2020kbt,FileviezPerez:2021ycy}). Owing to the addition of right-handed neutrinos~(RHNs) to the SM fermion sector, there exists a common bridge between the neutrino mass generation mechanism and a process which can explain the baryon asymmetry through the process of leptogenesis. This attractive idea of baryogenesis via the process of leptogenesis was first pointed out in Ref.~\cite{Fukugita:1986hr}. This idea is realized when the RHNs, the key ingredient of the {\it type-I seesaw} mechanism~\cite{Minkowski:1977sc,Mohapatra:1979ia,GellMann:1980vs}, can undergo CP violating decay and create ample amount of lepton asymmetry which further can account for the excess of matter over antimatter. This aesthetic connection has also led to a significant amount of work in the leptogenesis, in which the baryon number ($B$) is generated by the out-of-equilibrium lepton number ($L$) violating decays (for more detail please see \cite{Minkowski:1977sc,Mohapatra:1979ia,GellMann:1980vs,Yanagida:1979as,Ellis:1992nq,Rubakov:1996vz}) of heavy Majorana neutrinos $N_i$ to the SM leptons and Higgs. Furthermore, it was argued that the excess in $L$ will then be converted into the observed excess in $B$ by means of $B + L$ violating sphaleron interactions \cite{PhysRevD.49.6394,DOnofrio:2012phz}, which remain in thermal equilibrium above the critical temperature $T_c \sim 149$~GeV. The excess of matter over antimatter has been evidenced by many experimental observations. This excess of matter in the present Universe is expressed by a quantity called baryon to photon ratio ($\eta_B$), which has been estimated by the recent Planck satellite experiments as \cite{Aghanim:2018eyx,ade2016planck} 
\begin{equation*}
\eta_{B}=\frac{n_{B}-n_{\overline{B}}}{n_{\gamma}}=(6.02 - 6.18)\times 10^{-10}\ .
\end{equation*} 
where $n_B$, $n_{\bar B}$ and $n_\gamma$ are respectively the number densities of baryons, anti-baryons, and photons.

In addition to the traditional type-I seesaw mechanism, the idea of having leptogenesis can also be extended to any model that involves the presence of RHNs in order to explain the tiny neutrino mass. Possible realizations of such models explaining the matter-antimatter asymmetry have been studied for decades, starting with an attributed investigation in the context of grand unified theories \cite{Hambye:2001eu,Hambye:2004jf,Iso:2010mv,Okada:2012fs,Hambye:2016sby,Bodeker:2020ghk}. In our setup, the introduction of RHNs with non-trivial gauge charges will induce different HDOs that can essentially reduce to an effective type-I seesaw setup, along with other more general outcomes.
The fundamental origin of these seesaw operators is nothing but the $D=5$ Weinberg operator~\cite{PhysRevLett.43.1566}. From beyond the SM (BSM) perspective, we need to explore new avenues of neutrino mass generation. The most popular, seesaw doctrine is discussed beyond redemption and on the face of it, has no new insight to offer (probing them in the colliders is obviously one task at hand). 
Another plausible explanation is scotogenic neutrino models where loop diagrams at different orders can arrange for the smallness of the neutrino mass~\cite{Borah:2017dqx,Cai:2017jrq}. 
These two approaches can be combined into a general form when we directly write higher-dimensional operators 
(HDOs) to generate the neutrino mass terms, and the power of HDOs will be used to generate neutrino mass so tiny. 
These HDOs cannot be completely arbitrary as that will require a huge number of those operators. To get a sense of what kind of operators we should introduce, we take inspiration from a Frogatt-Nielsen-type~\cite{Froggatt:1978nt} flavor model to build an $\mathrm{U}(1)$ extended BSM theory and the guiding principle of vanishing gauge anomalies determines the form of the HDOs. In Ref.~\cite{Bonnefoy:2019lsn}, UV-complete Frogatt-Nielsen-like models are constructed to address fermion mass and mixing hierarchies. A full UV-complete realization of our HDOs is beyond the scope of this work. 
How the HDOs provide tiny masses to the SM neutrinos and also satisfy other measured neutrino oscillation parameters were already discussed in a companion work~\cite{Choudhury:2020cpm}. Similar, anomaly-free $\mathrm{U}(1)$ extensions 
and their interesting phenomenology is discussed in many different contexts in the literature~\cite{Appelquist:2002mw,Langacker:2008yv,Basso:2008iv,Coriano:2014mpa,Das:2016zue,Ekstedt:2016wyi,Ekstedt:2017tbo,Jana:2019mez,Sadhukhan:2020etu,Das:2021esm}.

In a minimalistic scheme where the seesaw mechanism is embedded with two RHNs~\cite{Ibarra:2003up}, one of the three active neutrinos remains massless. 
Leptogenesis in such an economical version of the type-I seesaw scenario has gained lots of attention and has widely been explored by the authors in a recent Ref.~\cite{Xing:2020ald}, starting from the electroweak to the grand unification scale.  In a type-I-like scheme with three RHNs, it is an appealing choice to place one of the RHN states (preferably the lightest one) at the lowest strata of the mass hierarchy. This choice can also motivate the inception of this model from the context of accommodating a potential dark matter candidate. With that arrangement, the theory becomes viable for explaining many of the key puzzles of the SM. In the present analysis, we propose a BSM framework where the lightest RHN state can act as a non-thermal DM candidate, while the heavier two participate in the process of leptogenesis in order to account for the observed BAU.

Depending on the temperature regime where the leptogenesis is assumed to take place, there exist relevant approaches to realize the process of baryogenesis through leptogenesis. 
Based on the temperature regime where leptogenesis works, one has to be careful while computing the lepton asymmetry, as this asymmetry essentially becomes dependent on the lepton flavors. 
For a detailed analysis of the flavor effects on thermal leptogenesis, one may look into \cite{Giudice:2003jh}. 
At the classical seesaw scale (typically about $10^{13}$ GeV), which is realized as a standard hierarchical leptogenesis scenario~\cite{Fukugita:1986hr} (also termed as {\it vanilla} leptogenesis), the SM neutrino mass limits (sub-eV) require experimentally unreachable mass scales for the heavy RHNs. 
Moreover, the required CP-violation in the theory can remain completely decoupled from the low-energy CP-violating source of the leptonic sector of the SM. This leads to the difficulty of experimentally probing vanilla leptogenesis. 
However, the scale of leptogenesis can be brought down to the TeV scale if a quasi-degenerate spectrum is assumed for the RHNs~\cite{Pilaftsis:2003gt,Deppisch:2010fr}. 
Most significantly this quasi-degeneracy among the RHN mass eigenstates is the manifestation of a mechanism that facilitates the low-scale leptogenesis, a framework termed as ``{\it resonant leptogenesis}''. 
Owing to the presence of the mentioned mass-degeneracy, the amount of the lepton asymmetry gets a resonant enhancement~(for detail please see \cite{Pilaftsis:2003gt,Deppisch:2010fr}). Any low-scale seesaw models can, in principle, explain the BAU through leptogenesis by the CP-violating decay of a low-scale (typically TeV-scale) RHNs. 
As a bonus, these low-scale seesaw models also set enough motivation for validating them in the context of the TeV-scale RHN production at the colliders. This is what constitutes the primary motivation of the present analysis.

The RHNs do not couple to the SM fields (very tiny couplings arise due to their mixing with the SM neutrinos) as they are singlet under the SM gauge group. This usually makes the production of the RHNs at the LHC very suppressed. In our model, however, the RHNs are charged under the new $\mathrm{U}(1)$ group to make the theory anomaly-free. This opens up an interesting possibility to produce the RHNs in pairs through the decay of the heavy $Z'$, the gauge boson of the new $\mathrm{U}(1)$, which can be copiously produced in $pp$ collisions at the LHC. Apart from the decay of $Z'$ to RHNs, it can also decay to the SM fields and leads to stringent constraints on the model parameters. The dominant constraints come from the dilepton resonance search data. In our analysis, we satisfy all the relevant constraints and find out the region of allowed parameter space that can be probed at the high luminosity LHC (HL-LHC) runs. In particular, we consider the same-sign dilepton signature that originates from the decay of the RHN pair. As this signal is almost background-free, we have performed a simplified analysis using this smoking-gun same-sign dilepton signature present in our model. In terms of the parameters, the leptogenesis and collider sectors are mostly connected through the leptonic Yukawa couplings. Although leptogenesis is sensitive to these Yukawas, they are not very collider sensitive as they appear in the branching ratios (BRs) of the RHNs. 
%%%%%%%%%%%

The rest of this article is framed as follows. In Section~\ref{sec:model}, we detail the construction of our model and the relevant mechanism associated with the neutrino mass generation. A brief discussion on the leptogenesis mechanism has been provided in Section~\ref{sec:lepto}. Section~\ref{sec:analysis} is kept for the numerical analysis required for fitting the neutrino oscillation data in order to construct the Yukawa coupling matrix for the calculation of lepton asymmetry. In Section~\ref{sec:collider}, we investigate on the collider prospects of this model. Finally, we conclude in Section~\ref{sec:conclusion}.

\section{The model: anomaly-free $\mathrm{U}(1)$ extensions}
\label{sec:model}
Here we explore an $\mathrm{U}(1)$ extended model with higher dimensional operators. In this setup, the gauge 
sector of the SM (i.e. $\mathrm{SU}(3)_c\times\mathrm{SU}(2)_L\times\mathrm{U}(1)_Y$) is extended by an additional $\mathrm{U}(1)_z$ symmetry, with gauge coupling $g_z$. One new gauge boson appears in the theory, which we introduce as a gauge eigenstate $Z^{\prime}_{\mu}$.  We also assume that the
SM fields are charged under new $\mathrm{U}(1)_z$ and the
corresponding quantum numbers ensure gauge anomaly cancellation.
Three RHNs, $N_i$ with $i= 1,2,3$ are added with non-trivial $\mathrm{U}(1)_z$ charges. 
Two of the RHNs are in 
the TeV range and one of them is ultra-light with mass in the KeV range. These extra neutrinos will 
eventually pave the way to achieve resonant leptogenesis through RHN decay at the TeV scale. 
Above the Electroweak Symmetry Breaking (EWSB) scale,
the $\mathrm{U}(1)_z$ is broken by beyond the SM singlets $S_i$ carrying 
charges $z_{S_i}$, which are such that trilinear terms in the scalar potential are not admissible.

\subsection{$\mathrm{U}(1)_z$ charges: anomaly cancellation}
In the fermionic sector, we assume that  
the $\mathrm{U}(1)_z$ charges for the SM fermions are family independent. The concerned fermions are 
assigned new gauge charges as $z_Q$ for the quark doublets, as $z_L$ for the lepton doublets. These charges for 
the right-handed up-type and down-type quarks are $z_u$ and $z_d$ respectively, while $z_e$ is the charge for
the right-handed charged leptons. For the newly added fermions, charges assigned to three RHN fields $N_i$ are 
$z_i$, all of which are not necessarily be same. In the scalar sector, the SM Higgs has a $\mathrm{U}(1)_z$ charge $z_H$, while the 
BSM scalars $S_i$ have charges $z_{S_i}$. The triangle gauge anomalies that appear due to the introduction of a new gauge group
$\mathrm{U}(1)_z$ can be analytically defined as ${\cal A} \equiv {\rm Tr}_L(T_a T_b T_c) - {\rm Tr}_R(T_a T_b T_c)$, 
where $T_a$ are the symmetry generators of the concerned symmetry group and the traces are taken over all left-handed and
right-handed fermions.  

Within the SM, the Yukawa terms (that give rise to the fermionic mass terms after EWSB) for the charged fermions
can be written as,
\begin{equation}
  {\cal L}_{\rm Yuk.} = y^{u}_{ij} \bar Q_{Li} u_{Rj} \widetilde H 
                    + y^{d}_{ij} \bar Q_{Li} d_{Rj} H 
                    + y^{e}_{ij} \bar L_{Li} e_{Rj} H + \textrm{H.c.} \ ,
\end{equation}
where $\widetilde H = i \sigma_2 H^*$.

Satisfying the gauge symmetry after its extension beyond the SM, these Yukawa terms for the charged 
SM fermions relate the gauge charges as,
\begin{equation}
z_H = z_L - z_e = z_Q - z_d = z_u - z_Q \ .
\end{equation}
Now we impose the constraints that come from the cancellation of different triangle anomalies
and subsequent relations between $\mathrm{U}(1)_z$ charges of those are tabulated in Ref.~\cite{Choudhury:2020cpm}. 
From the anomaly condition arising from $\left[\mathrm{SU}(3)_c\right]^2 \mathrm{U}(1)_z$ either $z_{u(d)}$ can be 
written as a combination of $z_Q$ and $z_{d(u)}$. From $\left[\mathrm{SU}(2)\right]^2 \mathrm{U}(1)_z$, $z_L$ can be expressed in terms of $z_Q$, while $z_e$ is a combination of $z_Q, z_{u(d)}$.  
Therefore, to express the $\mathrm{U}(1)_z$ charges of the SM fields, just two parameters, say $z_u$ and $z_Q$ can be consistently taken. It is worthwhile to note that in any $\mathrm{U}(1)$ extended model, without any loss of generality, 
one combination of charges can always be taken to be unity. In this model,
the combination $z_u - 4 z_Q = 1$ is chosen to finally render $z_Q$ to
be the only free parameter, so that $\mathrm{U}(1)_z$ charges of all the SM fields are expressed in terms of $z_Q$. 
It is worth mentioning that the mixed gauge-gravity anomaly ($R^2 \mathrm{U}(1)_z$) cannot impose any new constraint.
The corresponding charge assignments are displayed in Table.~\ref{tab:charges}.

Among the BSM particles, only the RHNs contribute to different gauge anomalies. 
From the $\lt[\mathrm{U}(1)_z\rt]^3$ anomaly, the charges $z_i$ for the $N_{iR}$ fields will require to satisfy,
\[
   \sum_{i=1}^3 z_i^3 = 3 \, (z_u - 4 z_Q)^3\ ,
\] 
which with our earlier consideration will reduce to,
\[
   \sum_i z_i^3 = 3 \ ,
\]
so that the anomaly quoted above vanishes. Barring the irrational and complex values, 
if we stick to integer values, the trivial choice that comes to our mind is $z_i =
1$, which was explored in detail in a different
context~\cite{Appelquist:2002mw}. This choice is not an optimal one from the context of 
neutrino mass generation. Consequently, we choose the next simplest integer option, assigning charges 
as $z_{1,2} = 4, \ z_3 = -5$. 
We cannot assign any particular $\mathrm{U}(1)_z$ charges to the BSM scalars from the anomaly cancellation point of view. 
We keep them as free parameters for now, which will be fixed when we discuss the Yukawa interactions in the context of BSM sector in the next subsection. 

\begin{table}
\centering
\vspace{0.5em}
\begin{tabular}{|c|ccc|l|}
\hline
Particle Content &$\mathrm{SU}(3)_{c}$&$\mathrm{SU}(2)_{L}$&$\mathrm{U}(1)_{Y}$&$\mathrm{U}(1)_{z}$\\
\hline
$q_{L}$&3&2&$1/6$&$z_{Q}$\\
$u_{R}$&3&1&$2/3$&$1 + 4 z_{Q}$\\
$d_{R}$&3&1&$-1/3$&$-1 - 2z_{Q}$\\
%\hline
$\ell_{L}$&1&2&$-1/2$&$-3z_{Q}$\\
$e_{R}$&1&1&$-1$&$-1 -6z_{Q}$\\
\hline
$N_{1R}$&1&1&0&$4$\\
$ N_{2R}$&1&1&0&$4$ \\
$N_{3R}$&1&1&0&$-5$\\
\hline
$H$&1&2&$1/2$&$1 + 3z_{Q}$\\
$S_1$&1&1&0&$z_{S_1}$\\
$S_2$&1&1&0&$z_{S_2}$\\
\hline
\end{tabular}
\caption{The charge assignments for the fermions and scalars of the model.}\label{tab:charges}
\end{table}

\subsection{Fermionic sector: neutrino mass and interaction}
One of our primary goals here is to explain Leptogenesis and consequent baryon over-abundance through the TeV-scale
RHN decays. In that direction, we have included extra neutral fermions and we need at least two of
them, which will play their obvious role in generating neutrino mass. Invoking three such RHNs is not only
enough to ensure the cancellation of all possible
anomalies, but also leads to very interesting phenomenological consequences. 
Non-trivial assignment of $\mathrm{U}(1)_z$ charges for the RHNs give rise to different types of BSM scalar assisted 
HDOs in this effective theory. Analogous HDOs can be written for the SM fermions, 
but they will be sub-dominant compared to the usual Yukawa terms. 

Instead of being a fundamental theory, where dimension-4 gauge invariant Yukawa operators are present as in the SM, 
the model discussed here is only an effective low-energy version of some more fundamental theory. 
The fully renormalizable model is valid at some scale $\Lambda$ or higher. 

The HDOs of the effective field theory
take the form,

\begin{equation}
\begin{array}{rcl}
{\cal L}_{\nu } &= & \displaystyle {\cal L}_{\rm Dirac} + {\cal L}_{\rm Wein.};
\\[2ex]
{\cal L}_{\rm Dirac} & = & \displaystyle \sum_{i= 1}^3 \sum_{\alpha = 1}^2 
        y_{\alpha i} N_{\alpha R} \bar L_{iL} \widetilde H \; 
            \frac{S_1^{a_1} \, S_2^{a_2}}{\Lambda^{|a_1|+ |a_2|}}
          + \sum_{i= 1}^3   y_{3i} N_{3 R} \bar L_{iL}  \widetilde H  \;
            \frac{S_1^{a_3} \, S_2^{a_4}}{\Lambda^{|a_3|+ |a_4|}} + \textrm{H.c.};
\\[2ex]
{\cal L}_{\rm Wein.} & = & \displaystyle \sum_{i,j= 1}^3 
        s_{i j} \overline{L^c_{iL}} L_{jL} H H 
            \frac{S_1^{b_1} \, S_2^{b_2}}{\Lambda^{|b_1|+ |b_2|+1}}
        + \sum_{\alpha,\beta= 1}^2  s_{\alpha\beta} \overline{N^c_{\alpha R}} N_{\beta R} 
            \frac{S_1^{b_3} \, S_2^{b_4}}{\Lambda^{|b_3|+ |b_4|-1}}
\\[2ex]
& + & \displaystyle
    \sum_{\alpha=1}^2  s_{\alpha3} \overline{N^c_{\alpha R}} N_{3 R} 
            \frac{S_1^{b_5} \, S_2^{b_6}}{\Lambda^{|b_5|+ |b_6|-1}}
        +  s_{33} \overline{N^c_{3R}} N_{3 R} 
            \frac{S_1^{b_7} \, S_2^{b_8}}{\Lambda^{|b_7|+ |b_8|-1}} + \textrm{H.c.},
\end{array}
\label{sc-terms}
\end{equation}
where the couplings $y_{i \alpha}, y_{i3}, s_{ij}, s_{\alpha\beta},
s_{\alpha3}$ and $s_{33}$ are dimensionless and to satisfy the gauge symmetry along with the 
conditions arising from anomaly cancellation the exponents get related as,
\begin{equation}\label{scalar:constraint}
\begin{array}{rcl c rcl}
z_{S_1} a_1 + z_{S_2} a_2 & = & -3  & \qquad \quad &
z_{S_1} a_3 + z_{S_2} a_4 & = & 6
\\[1ex]
z_{S_1} b_1 + z_{S_2} b_2 & = & -2  & \qquad \quad &
z_{S_1} b_3 + z_{S_2} b_4 & = & -8
\\[1ex]
z_{S_1} b_5 + z_{S_2} b_6 & = & 1  & \qquad \quad &
z_{S_1} b_7 + z_{S_2} b_8 & = & 10 \ .
\end{array}
\end{equation}
We are going to stick to the integer solutions for the exponents
to avoid the emergence of nonlocal
operators. Negative values for the exponents are permissible as, if they appear, they can be understood as
positive powers of $S_{1,2}^{*}$. Possible origin of these operators in a UV-complete theory, integrating out 
heavy ($ \ge \Lambda$) vectorlike fermions, and the relevant complexities are discussed in Ref.~\cite{Choudhury:2020cpm}. For $x_i$ indicating the vacuum expectation values of the new scalars $S_{1,2}$, $x_i$ must be in the range $\Lambda > x_i \gapp \Lambda/(4\pi)$~\cite{Choudhury:2020cpm} while obtaining the effective terms in Eq. (\ref{sc-terms}) to avoid including the loop induced HDOs. As the terms included in ${\cal L}_{\rm Wein.}$ violate
lepton-number, corresponding mass terms that we obtain after $\mathrm{U}(1)_z$
symmetry breaking can be thought of as Majorana masses. The appearance of Majorana mass in this setup is argued in Ref.~\cite{Choudhury:2020cpm}.

From Eq.~\eqref{scalar:constraint} above, one can make a number of choices for the $\mathrm{U}(1)_z$ gauge charges of the 
scalars $S_i$, to find out the exponents $a_i$. 
As a very specific case, the assignment 
\beq
    z_{S_1} = -3/4 \ , \qquad z_{S_2} = -4 ,
\eeq
is taken, which may lead to some interesting phenomenology. 
This is not a special one and similar qualitative results will be obtained 
for different other choices. This choice results in relatively unsuppressed masses in the $N_{1,2}$ sub-sector,
while terms involving $N_{3R}$ are suppressed by higher orders of $\Lambda$. After the symmetry breaking and expansion in terms of fields around the vacuum expectation values (VEVs), the terms in the first order in regard to least suppression,
\beq
\barr{rcl}
{\cal L}_{\rm Dirac} & = & \displaystyle \sum_{i= 1}^3 \sum_{\alpha = 1}^2 
        y_{\alpha i} \bar N_{\alpha R} L_{iL}  \widetilde H \; 
            \frac{S_1^{4} }{\Lambda^{4}}
          + \sum_{i= 1}^3   y_{3 i} \bar N_{3 R} L_{iL}  \widetilde H  \;
            \frac{S_1^{*8} }{\Lambda^{8}} + \textrm{H.c.}\ ,
\\[2ex]
{\cal L}_{\rm Wein.} & = & \displaystyle \sum_{i,j= 1}^3 
        s_{i j} \overline{L^c_{iL}} L_{jL} H H 
            \frac{S_1^{8} \, S_2^{*}}{\Lambda^{10}}
        + \sum_{\alpha,\beta= 1}^2  s_{\alpha\beta} \overline{N^c_{\alpha R}} N_{\beta R} 
            \frac{S_2^{2}}{\Lambda}
\\[2ex]
& + & \displaystyle
    \sum_{\alpha=1}^2  s_{\alpha3} \overline{N^c_{\alpha R}} N_{3 R} 
            \frac{S_1^{4} \, S_2^{*}}{\Lambda^{4}}
        +  s_{33} \overline{N^c_{3R}} N_{3 R} 
            \frac{S_1^{8} \, S_2^{*4}}{\Lambda^{11}} + \textrm{H.c.}\ .
\end{array}
 \label{nu_mass_our}
 \eeq
The terms corresponding to $s_{ij}$ and $s_{33}$ are negligible and therefore can be dropped hereafter. 
We restrict ourselves to HDOs of mass dimension 12 or less, to write the relevant terms as, 
\beq
\barr{rcl}
{\cal L}_{\nu}  &\approx & \dis {\cal L}^{(5)} + {\cal L}^{(8)} 
                                    + {\cal L}^{(12)} + \textrm{H.c.}
\\[2ex]
\dis {\cal L}^{(5)} & \equiv & \dis 
        \sum_{\alpha,\beta= 1}^2  s_{\alpha\beta} \overline{N^c_{\alpha R}} N_{\beta R} 
            \frac{x_2^{2}}{\Lambda},
\\[2ex]
\dis {\cal L}^{(8)} & \equiv & 
\dis \sum_{i= 1}^3  \sum_{\alpha = 1}^2 
        y_{\alpha i} \bar N_{\alpha R} L_{iL}  \widetilde H \; 
            \frac{x_1^{4} }{\Lambda^{4}}
    + \sum_{\alpha=1}^2  s_{\alpha3} \overline{N^c_{\alpha R}} N_{3 R} 
            \frac{x_1^{4} \, x_2^{*}}{\Lambda^{4}},
\\[2ex]
\dis {\cal L}^{(12)} & \equiv & \dis 
\sum_{i= 1}^3   y_{3 i} \bar N_{3 R} L_{iL}  \widetilde H  \;
            \frac{x_1^{*8} }{\Lambda^{8}} \ ,
\earr
\label{nu_mass_simpl}
\eeq
and this is the form that we would be working with henceforth.

The neutrino mass matrix can be constructed from the terms given in Eq.~\eqref{nu_mass_simpl}. It can be
represented in the $(\nu_j, N_1, N_2, N_3)$ basis, by 
\beq 
{\cal M}_\nu = \left( \barr{cc} 0_{3\times 3} & \mc{D}^T \\ {\cal D} & M_N
\earr \right) .  
\eeq 
Denoting $\xi \equiv x/\Lambda$, where $x$
is either of $x_{1,2}$, assuming that there is no large
hierarchy between the $x_i$, the matrices above have structures (in terms of Yukawa parameters)
%%%%%%
\beq \label{eq:mD}
{\cal D} \approx v \xi^4 \left(
\begin{array}{ccc}
 y_{11}  & y_{12} & y_{13} \\
 y_{21}  & y_{22} & y_{23} \\
 y_{31} \xi^4 & y_{32} \xi^4 & y_{33} \xi^4 \\
\end{array}
\right)
 \qquad
M_N \sim \frac{x^2}{\Lambda} \, \left(
\begin{array}{ccc}
a_1 & 0 & s_{13} \xi^3  \\
0 & a_2 & s_{23} \xi^3 \\
s_{13} \xi^3 & s_{23} \xi^3 & 0 \\
\end{array}
\right).
\eeq
Introducing two free parameters, $a_1$ and $a_2$ in the $N_1$-$N_2$ sub-sector
of the mass matrix, without any loss of
generality, their values can be fixed so that the matrix will be diagonal. 

Due to moderate suppression of the form  {\em viz.} ${\cal
O}(x^2/\Lambda)$, two eigenvalues of $M_N$ are large which are fixed to be at the TeV scale. 
The heavier RHNs need to be quasi-degenerate to produce lepton asymmetry 
through resonant Leptogenesis.
These heavy neutrinos tend to be lighter than
the BSM gauge boson $Z'$, emerging from the new gauge group $\mathrm{U}(1)_z$. 

The third RHN is much lighter, with a mass of the order of $ \sim {\cal
O}(\xi^6 x^2/\Lambda)$. The lightest neutrino mixes minimally with the
heavier ones, with the mixing angles being ${\cal O}(\xi^4)$ suppressed.

\subsection{Bosonic sector: gauge bosons and scalars}
The gauge boson sector delves in the gauge boson mass generation through symmetry breaking, which in turn, generates masses of the scalars through different mixings. The symmetry breaking happens in two steps, first $\mathrm{U}(1)_z$ symmetry is broken and then we have EWSB. Both are broken by the VEVs, respectively of  the $S_i$ and $H$ fields.
\begin{equation}
\langle S_A \rangle \equiv \frac{x_i}{\sqrt{2}} \ , \qquad
\langle H \rangle \equiv \left(0 \quad \frac{v_h}{\sqrt{2}} \right)^T \ .
\end{equation}
We assume $x_i$ are of the same order, i.e., no hierarchy is
introduced between them. Furthermore, as a simplistic viewpoint, we
take all the VEVs to be real to avoid spontaneous $CP$-violation in the scalar sector.

%\\
After both of the symmetries are broken, the resultant
 heavy neutral gauge boson masses are given by 
%%%%
\begin{equation}
\label{eq:mass}
M_{Z,Z'}^2 = \dfrac{e^2 v_h^2 \cos^2 \theta}{\sin^2 2 \theta_W} 
       +\frac{g_z^2}{4}\left(z_H^2 v_h^2 + \sum_A z_{\chi_A}^2 x_A^2\right) 
              \sin^2 \theta 
        \mp \dfrac{e g_z z_H v_h^2}{2\sin 2 \theta_W} \sin 2\theta\ ,
\end{equation}
 where the $Z\leftrightarrow Z^\prime$ mixing angle $\theta$ is given by
\begin{equation}
\dfrac{4e z_Hg_z}{\sin 2 \theta_W} \, \cot 2 \theta 
     = \frac{g_z^2}{v_h^2}\Big(\sum_A z_{S_A}^2 x_A^2+z_H^2 v_h^2\Big) 
      - \dfrac{4 e^2}{\sin^2 2 \theta_W}\ .
\end{equation}
We need to keep the $Z$ mass fixed at the measured values, and in the allowed parameter space, the $Z^{\prime}$ 
masses (which is at multi-TeV range) and properties need to be probed. Interplay of previously introduced RHN and the $Z^{\prime}$ 
phenomenology is crucial in 
observing many important decay modes. 

The covariant derivative,
where the $Z^{\prime}$ is a part of it, used to write the RHN kinetic term which leads to $Z^{\prime} \to N_i N_i$ 
decays, eventually resulting in a same-sign dilepton final state.

In the scalar sector, we consider two scalar
fields $S_1$ and $S_2$, which would prove to be of particular
interest in the context of neutrino mass generation. If the 
corresponding $\mathrm{U}(1)_z$ charges $z_{S_1}$ and $z_{S_2}$ are not 
integral multiples of each other i.e. 
the ratio of their gauge charges will be different from simple integers. the most 
general form of the scalar potential invariant under $\mathrm{U}(1)_z$ is given 
by
\begin{equation}
V(S_1, S_2) = - \mu_1^2 S_1^\dagger S_1 - \mu_2^2 S_2^\dagger S_2 
                  + \frac{\lambda_1}{2}\,  (S_1^\dagger S_1)^2
                  + \frac{\lambda_2}{2}\,  (S_2^\dagger S_2)^2
              + \lambda_{12}\, (S_1^\dagger S_1)\, (S_2^\dagger S_2) \ .
\label{2chi_pot}
\end{equation}
We start with an assumption that the $S$-sector
is essentially decoupled from the SM Higgs sector, which gauge charge assignments can ensure. 
At higher energy scales than the EWSB scale, this is a justifiable approximation, except for some decays of the new 
scalars $S_i$ into the SM Higgs, which requires some coupling (mixing) between them. With the decoupling, the potential is complete with the terms written in Eq.~\eqref{2chi_pot}. 
Writing the $S_i$ fields in their post-symmetry breaking form

with $\xi_{1,2}, \rho_{1,2}$ as real fields, the massless 
pseudoscalar (along with the mixing angle $\gamma_A$) appears as a combination:
\begin{equation}
    A = \rho_1 \, \sin \gamma_A - \rho_2 \, \cos\gamma_A \ , 
    \qquad \tan \gamma_A = \frac{z_{S_2} x_2}{z_{S_1} x_1},
\end{equation}
with the orthogonal combination being absorbed as a Goldstone boson of the $Z'$. 

The mass-squared matrix for the two CP-even scalars
$\xi_{1,2}$ reads
\[
    M^2_{\xi_i} = \left( \begin{array}{ccc} 
                 \lambda_1 x_1^2 & \quad & \lambda_{12} x_1 x_2 \\[1ex]
                 \lambda_{12} x_1 x_2  & \quad & \lambda_1 x_1^2
                 \end{array} \right),
\]
leading to mass eigenstates $H_{1,2}$ with the corresponding masses defined as

\begin{equation}
M^2_{H_1,H_2} = \frac{1}{2} \left[ \lambda_1 x_1^2 + \lambda_2 x_2^2
             \pm |\lambda_1 x_1^2 - \lambda_2 x_2^2| \;
     {\rm sec}(2 \alpha_\chi) \right],
\end{equation}
where the mixing angle is written as
\begin{equation}
\qquad \tan (2\alpha_\chi) = 
     \frac{2 \lambda_{12} x_1 x_2}{\lambda_1 x_1^2 - \lambda_2 x_2^2}.
\end{equation}
The phenomenology of these BSM scalars can be interesting only if we can couple them to the SM Higgs sector, which is beyond the scope of this work.

%============================================================================
\section{Baryogenesis through resonant leptogenesis}\label{sec:lepto}

In order to evaluate the viability of low-scale leptogenesis, it is essential to consider the effects of all the lepton flavors while calculating the lepton asymmetry (details can be found in \cite{Dev:2017trv,Davidson:2007xu}).
This is because, at such temperature regimes, all the lepton flavors remain in equilibrium and hence can act non-identically. As it is known that different lepton
flavors are distinguished by their Yukawa couplings $Y_\nu^\ell$ ($\ell = e,\mu~\text{and}~\tau$). We learn that the $\ell$-th lepton flavor becomes distinguishable when $Y_\nu^\ell$ related interactions enter into equilibrium.  The corresponding $Y_\nu^\ell$ related
interaction rates are then compared with the Hubble expansion rate of the Universe, which let us know that the $Y_\nu^e, Y_\nu^\mu,\; \text{and} \;Y_\nu^\tau$ related
interactions are in equilibrium below certain temperature regimes. Moreover, the flavor effects are relevant to be taken into account as they also have an influence on the washout effect. The lepton doublets actively involve in the washout processes as the initial states, so one needs to know which flavors are distinguishable before determining the washout interaction rates. Such interesting flavor effects should in principle be considered in executing the baryogenesis-through-leptogenesis idea, if the RHN mass~(after the seesaw) scale is below $10^{12}$ ~GeV.

Apart from the flavor effects, there exist some more interesting features of such low-scale leptogenesis. In what follows, the  yield of lepton asymmetry is larger in a low-scale leptogenesis in comparison to that of a high scale thermal leptogenesis scenario. Traditionally, a resonant enhancement in the lepton asymmetry is seen when the RHN mass spectrum has a certain amount of degeneracy comparable to their decay width as mentioned in Ref.~\cite{Pilaftsis:2003gt}. For a detailed field theoretical justification supporting this fact, one may look into Refs.~\cite{Pilaftsis:1997jf,Flanz:1996fb,Xing:2006ms}. This complementarity between this resonance enhancement and degeneracy of the RHN masses can be understood from the following relations
\begin{equation}
M_{Ni} - M_{Nj} \sim \frac{\Gamma_{N_{i,j}}}{2},\;\;\; \frac{\text{Im} \left(Y_\nu^\dagger Y_\nu\right)^2_{ij}}{ \left(Y_\nu^\dagger Y_\nu\right)_{ii} \left(Y_\nu^\dagger Y_\nu\right)_{jj}}\sim1.
\end{equation}
Exactly satisfying the first of the above conditions can even lead to a lepton asymmetry of $\mathcal{O}(1)$. In the present analysis, we have well taken care of this requirement through a proper choice of model parameter space with the Yukawa couplings which govern the tiny neutrino mass through the HDOs.

The aforementioned quasi-degenerate spectrum in the RHN sector as required for resonant leptogenesis can be achieved in many ways. Without imposing any extra flavor symmetry in the RHN sector, one can achieve it by the interplay of the free parameters of the model. Although this way of doing is fine-tuned and looks less appealing, we present it in the main text to keep our result generic. In Appendix~\ref{app:flavsym}, we show an explicit example of discrete $S_3$ flavor symmetry (and also $N_1\leftrightarrow N_2$ exchange symmetry) imposed in the RHN sector where the quasi-degenerate RHN spectrum can be naturally achieved. We find that with the imposition of the additional symmetry, the parameter space will become more constrained. In alternative $\mathrm{U}(1)_{B-L}$ model (similar RHN charge assignment under new $\mathrm{U}(1)$ as ours) in Ref.~\cite{Asai:2020xnz}, the quasi-degenerate RHNs are obtained imposing a $Z_2$ symmetry.

As pointed out in~\cite{Fukugita:1986hr}, the out-of-equilibrium and CP-violating decay of heavy Majorana neutrinos provides a natural way to create the required lepton asymmetry. The asymmetry generated by the decay of the RHN (in the heavy sector) into a lepton and a Higgs is given by, 
    \begin{equation}
    \epsilon_{N_{k}}^\ell = -\sum \frac{\Gamma(N_{k} \rightarrow L_{\ell}+H^{+} , \nu_\ell+ H^0)-\Gamma(N_{k} \rightarrow L_{\ell}+H^-, \nu_\ell^c +H^{0^*})}
    {\Gamma(N_{k} \rightarrow L_{\ell}+H^{+} , \nu_\ell +H^0)+\Gamma(N_{k} \rightarrow L_{\ell}+H^{-}, \nu_\ell^c +H^{0^*})}
   \end{equation} 
Basically, CP-asymmetry is a measure of the difference in decay widths of $N_k$ to a process and its conjugate process. At the tree level, these two
are the same giving rise to vanishing CP-asymmetry. Therefore, we have to investigate with higher-order terms to obtain non-zero CP-asymmetry. Taking into account the one-loop vertex and self-energy diagrams, it is found that non-zero CP-asymmetry arises due to the interference between the tree level and the one-loop diagrams.

 Being one of the most consistent cosmological consequences of the type-I seesaw, leptogenesis demands the decay of the RHNs of mass around $\sim 10^{9 - 12}$~GeV which is difficult to reach the detectable regime of the colliders. Keeping that in mind, we have kept our motivation for TeV scale leptogenesis, as mentioned in the introduction. It is to note that with TeV scale RHNs, one can hope for an enhanced lepton asymmetry by considering the RH neutrinos $N_1$ and $N_2$ to be highly degenerate rather than exactly degenerate. It is clear from the asymmetry formula in Eq.~\eqref{eq:asymmetry} that exactly degenerate $N_1$ and $N_2$  would give no lepton asymmetry and hence we require non-zero mass splitting~(albeit very tiny).  As mentioned earlier, flavor-dependent effects of leptogenesis are relevant at low enough temperatures (set by the RHN mass) such that at least one charged lepton flavor is in thermal equilibrium. When this condition is met, flavor-dependent effects are not avoidable as the efficiency factors differ significantly for the distinguishable flavors. The relevant expression for the lepton asymmetry parameter, which is obtained by taking the effects of all the lepton flavors is given by~\cite{Deppisch:2010fr,Bambhaniya:2016rbb},
 \begin{equation}\label{eq:asymmetry}
    \epsilon_{i \ell} = \sum_{j \neq i} \frac{\text{Im}[Y_{\nu_{i \ell}}Y_{\nu_{j \ell}}^{*}(Y_{\nu}Y_{\nu}^\dagger)_{ij}]+\frac{M_i}{M_j}\text{Im}[Y_{\nu_{i \ell}} Y_{\nu_{j \ell}}^{*}(Y_{\nu}Y_{\nu}^\dagger)_{ji}]}{(Y_{\nu}Y_{\nu}^{\dagger})_{ii}(Y_{\nu}Y_{\nu}^{\dagger})_{jj}} f_{ij}^{mix}   
   \end{equation}
with the regulator given by,
\begin{equation*}
f_{ij}^{mix} = \frac{(M_{i}^2 - M_j^2)M_{i}\Gamma_{j}}{(M_{i}^2 - M_j^2)^2 + M_i^2 \Gamma_j^2}  
\end{equation*}
 with $\Gamma_i = \frac{M_i}{8\pi}(Y_\nu Y_\nu^\dagger)_{ii}$ as the tree level heavy-neutrino decay width. There is also a similar contribution $\epsilon_{il}^{osc}$ to the CP-asymmetry from RHN oscillation~\cite{Kartavtsev:2015vto,Dev:2015wpa}. Its expression is given by Eq.~\eqref{eq:asymmetry} with the replacement $f_{ij}^{mix} \rightarrow f_{ij}^{osc}$, where 
\begin{equation*}
f_{ij}^{osc} = \frac{(M_{i}^2 - M_j^2)M_{i}\Gamma_{j}}{(M_{i}^2 - M_j^2)^2 + (M_i \Gamma_i +M_j \Gamma_j)^2 \frac{\text{det}[\text{Re}(Y_\nu Y_\nu^\dagger)]}{(Y_\nu Y_\nu^\dagger)_{ii}(Y_\nu Y_\nu^\dagger)_{ii}}}  
\end{equation*}

It is evident from the above expression in Eq.~\eqref{eq:asymmetry} that, in order to yield a non-zero lepton asymmetry one has to have the Yukawa couplings ($Y_\nu$) as complex quantities. This can be made possible under the consideration of low energy CP-violation which  enters the PMNS matrix in the form of the Dirac and Majorana phases. This Yukawa coupling matrix can be obtained from the Dirac mass matrix (here, $\cal{D}$) in a basis such that the RHN mass matrix is diagonal and real in order to consider the decays of physical RHN states to the SM leptons and Higgs boson.  This can be achieved by writing $Y_\nu = Y_\nu^\prime U_R$, where $U_R$ serves as the diagonalizing matrix of the heavy RHN mass matrix ($M_N$). So one can simply extract $Y_\nu$ by writing $(\mathcal{D}/v)$ where $v$ is the VEV of the SM Higgs. 

Once we estimate the lepton asymmetry one can cast the analytically approximated expression for the baryon to photon ratio \cite{Pilaftsis:2003gt} as,
 \begin{equation}\label{eq:bau}
 \eta_B \simeq -3\times 10^{-2} \sum_{\ell,i}\frac{\epsilon_{i \ell}}{K_\ell^{\text{eff}}\text{min}(z_c,z_\ell)}
\end{equation}  
where, $z_c = \frac{M_N}{T_c}$, $T_c \sim 149$ GeV \cite{Pilaftsis:2003gt}, the critical temperature at which sphalerons read off. Here, $z_\ell \simeq 1.25 \text{Log}(25 K_\ell^{\text{eff}})$ and $K_\ell^{\text{eff}} = \kappa_\ell \sum_{i} K_i B_{i \ell}$ , with $K_i = \Gamma_i /H_N$, the washout factor. $H_N$ is $1.66 \sqrt{g^*}M_N^2/M_{\text{Pl}}$ is the Hubble expansion rate at temperature $\sim M_N$ with $g^* \simeq 106.75$ and $M_{\rm Pl} = 1.22 \times 10^{19}$~GeV. $B_{i \ell}$'s are the branching ratios of the $N_i$ decay to leptons of $l$th flavor : $B_{i \ell} = \frac{|Y_{\nu_{i \ell}}|^2}{(Y_{\nu}Y_{\nu}^{\dagger})_{ii}}$.
Including the RIS~(Real Intermediate State) subtracted collision terms one can write the factor $\kappa$ from the Ref.~\cite{Deppisch:2010fr},
\begin{equation}
\kappa_\ell = 2 \sum_{i,j j \neq i} \frac{\text{Re}[Y_{\nu i \ell}Y_{\nu j \ell}^* (YY^\dagger)_{ij}]+ \text{Im}[(Y_{\nu i \ell} Y_{\nu j \ell}^*)^2]}{\text{Re}[(Y^\dagger Y)_{\ell \ell} \{(Y Y^\dagger)_{ii} + (Y Y^\dagger)_{jj}\}]}\times\left(1-2i \frac{M_i-M_j}{\Gamma_i + \Gamma_j}\right)^{-1}.
\end{equation} 
 where, $Y_{\nu}$ is the Dirac Yukawa coupling matrix as mentioned earlier with $l$ as the lepton flavor index. It is worth noting that, during the calculation of RIS contribution since only the diagonal terms are considered in the sum, $\kappa_l$ can take its maximum value and hence we can have $\kappa_\ell = 1 + O(\delta_\ell^2)$. As seen from the expression Eq.~\eqref{eq:asymmetry}, the lepton asymmetry is dependent on the elements of the Dirac Yukawa coupling matrix. Therefore it can be said that the same set of model parameters that are supposed to yield correct neutrino phenomenology may also give rise to an expected order of lepton asymmetry, to finally reproduce the observed BAU. In the following section, we present the methodology of extracting the Yukawa parameters where the global fit oscillation parameters are used as primary inputs.

%==============================================
 \section{Numerical analysis and results}\label{sec:analysis}
 \subsection{Fitting with neutrino oscillation data}
 The mismatch between $\nu_\ell$ and $\nu_i$ can be explained by the well-known Pontecorvo-Maki-Nakagawa-Sakata (PMNS) matrix. In the simple case where the charged lepton mass matrix is diagonal which is true in this model, we can have $U_l = \mathbb{I}$. This leads us to write $U_{\text{PMNS}} = U_{\nu}$. Following this, the complete light neutrino mass matrix can be cast into as follows
\begin{equation}
m_{\nu}= U_{\text{PMNS}}m^{\text{diag}}_{\nu} U^T_{\text{PMNS}},
\label{nuoscillation}
\end{equation}
where the PMNS leptonic mixing matrix can be parametrized as
\begin{equation}
U_{\text{PMNS}}=\left(\begin{array}{ccc}
c_{12}c_{13}& s_{12}c_{13}& s_{13}e^{-i\delta}\\
-s_{12}c_{23}-c_{12}s_{23}s_{13}e^{i\delta}& c_{12}c_{23}-s_{12}s_{23}s_{13}e^{i\delta} & s_{23}c_{13} \\
s_{12}s_{23}-c_{12}c_{23}s_{13}e^{i\delta} & -c_{12}s_{23}-s_{12}c_{23}s_{13}e^{i\delta}& c_{23}c_{13}
\end{array}\right) U_{\text{M}}
\label{eq:PMNS}
\end{equation}
where $c_{ij} = \cos{\theta_{ij}}, \; s_{ij} = \sin{\theta_{ij}}$ and $\delta$ is the leptonic Dirac CP phase. The diagonal matrix $U_{\text{M}}=\text{diag}(1, e^{i\alpha}, e^{i(\beta+\delta)})$ contains the undetermined Majorana CP phases $\alpha, \beta$. The source of low energy CP violation enters in the above leptonic mixing matrix through the phases $\delta, \; \alpha, \; \text{and}\; \beta$. One can express the neutrino mass eigenvalues  with the help of two precisely measured solar and atmospheric mass splittings as $m^{\text{diag}}_{\nu} 
= \text{diag}(m_1, \sqrt{m^2_1+\Delta m_{21}^2},\sqrt{m_1^2+\Delta m_{31}^2})$ for normal hierarchy (NH) and $m^{\text{diag}}_{\nu} = \text{diag}(\sqrt{m_3^2+\Delta m_{23}^2-\Delta m_{21}^2}$, $\sqrt{m_3^2+\Delta m_{23}^2}, m_3)$ for inverted hierarchy (IH) as the diagonal mass matrix of the light neutrinos. The global fit oscillation parameters used to determine neutrino Yukawa coupling is presented in Table~\ref{tabdata}.
%==============================================
\begin{table}
\begin{center}
\begin{tabular}{|c|c|c|}
\hline
Parameters & Normal ordering & Inverted ordering  \\
\hline \hline
$\sin^2\theta_{23}$ & 0.433 - 0.609 & 0.436 - 0.610\\
\hline
$\sin^2\theta_{12}$ & 0.275 - 0.350 & 0.275 - 0.350 \\
\hline
 $\sin^2\theta_{13}$ & 0.02044 - 0.02435 & 0.02064 - 0.02457 \\
\hline
$\Delta m^2_{21}$ & (6.79 - 8.01)$\times 10^{-5}$ eV$^2$ & (6.79 - 8.01 ) $\times 10^{-5}$ eV$^2$\\
\hline
$\Delta m^2_{31}$ & (2.436 - 2.618)$\times 10^{-3}$ eV$^2$ & $-$(2.601 - 2.419)$\times 10^{-3}$ eV$^2$\\
\hline
$\delta_{CP}/ ^{\circ}$ &144 - 357 & 205 - 348 \\
\hline
\end{tabular}

\caption{Latest $3\sigma$ bounds on the oscillation parameters (with SK data) from Ref.~\cite{Esteban:2018azc}.}
\label{tabdata}
\end{center}
\end{table}
At present, the value of Dirac CP phase is ambiguous in the sense that, T2K experiment~\cite{Abe:2019vii} prefers $\delta_{\rm CP} \approx \pi/2$  whereas  NOvA~\cite{Acero:2019ksn} tells that $\delta$ can take CP conserving values. However, the global fit values for each hierarchy imply $\delta_{\rm CP} \approx -3\pi/4$ for NH and $\delta_{\rm CP} \approx -\pi/2$ for IH as also evident from the Table \ref{tabdata}. Taking care of the above neutrino oscillation constraints we extract the region of parameter space corresponding to the effective Yukawa couplings sitting in the HDOs and also present in the expression for the Dirac neutrino mass matrix (LHS Eq.) Eq.~\eqref{eq:mD} as we discuss in the following subsection.

Having set the stage, we perform a parameter scan using equations \ref{nuoscillation}. We vary all the oscillation parameters in their $3\sigma$ ranges tabulated in Table~\ref{tabdata} and numerically evaluate the neutrino Yukawa couplings ($Y^\nu_{i \ell}$) which carry the source of CP violation solely driven by the low energy CP phases.  It is to mention here that, while extracting the nine Yukawa parameters numerically we made an approximation among three of the Yukawa parameters ensuring that they do not disturb the neutrino observables. After we numerically evaluate these remaining Yukawa couplings, we feed them for the calculation of the lepton asymmetry parameter using Eq.~\eqref{eq:asymmetry}. Once we numerically estimate the lepton asymmetry ($\epsilon_{i \ell}$) we determine the BAU ($\eta_B$) through two approaches 1) using the analytically approximate solution (with the help of Eq.~\eqref{eq:bau}),  and  2) by exactly solving the coupled Boltzmann equations (BEQs) described in subsection \ref{sec:beq}. The purpose of choosing 1) is to reflect the exact values of the low energy CP phases from the requirement of satisfying $\eta_B$. The subsequent effect can be seen from the figures \ref{fig:epsilonNH} and \ref{fig:epsilonIH}. However, for having a clear vision on the evolution of lepton asymmetry through time it is advised to take the help of the exact solution of the BEQs. The corresponding realization is provided in Fig.~\ref{fig:NBL}. 
The flavor-dependent lepton asymmetries ($\epsilon_{i \ell}$) hence evaluated are then fed into the BEQs to get the final value of the lepton asymmetry which is further converted to the baryon asymmetry through the Eq.~\eqref{eq:etaL}. 

In order to investigate the process of leptogenesis, one has to calculate the CP asymmetry parameter which is very much dependent on the structure of the Dirac neutrino mass matrix and hence, depends upon the specific model under consideration. The leptogenesis phenomenon may be flavor dependent or independent according to the temperature regime. At extremely high temperatures (above $10^{12}$ GeV) flavor effects are unimportant and lepton asymmetry becomes flavor independent, but below this temperature one can not ignore the lepton flavor effects as the flavor-dependent interactions come into equilibrium \cite{Nardi:2006fx}.

For the desired regime of leptogenesis in this model, the degeneracy in terms of the mass splitting of the RH neutrinos is supposed to be of the order of their decay width which is followed by writing $M_1 = 1~\textrm{TeV},\; M_2 = M_1 - \Delta M$. $\Delta M$ being the desired mass splitting which is around $10^{-6}$ GeV for this analysis. It is however to note here that, this mass splitting can be naturally obtained in this model under tuning of some parameters which are involved in the construction of the heavy Majorana neutrino mass matrix (RHS of) Eq.~\eqref{eq:mD}. We have well taken care of the interplay between the degeneracy and the decay width of the RHNs ($\Gamma_{N_i}, i=1,2$), and notice that the resonant condition is absolutely essential to make this model viable for leptogenesis. 

%===============================================
\subsection{Lepton asymmetry: role of the CP phases}
After the announcement made by T2K \cite{Abe:2019vii}, it became an important job to find the source of high-energy CP violation in the low-energy CP violating phases. In principle, the low energy theory explaining the neutrino mass and mixing involves three CP phases namely, one Dirac ($\delta$) and two Majorana ($\alpha, \beta$) for three RHN model as described above in Eq.~\eqref{eq:PMNS}. Based on the fact that these phases take some CP-conserving or violating values, leads to interesting predictions in the process of baryogenesis. For some detailed work on such aesthetic connection please refer to \cite{Davidson:2002em,Pascoli:2003uh,Molinaro:2008rg,Moffat:2018smo}.  In this section, we try to examine the role played by the Dirac and Majorana CP phases in the present framework comprising of higher dimensional operators which offer the low energy Majorana neutrino mass. As evident, the lepton asymmetry parameter depends on the nature of the Yukawa couplings governing the neutrino mass generation along with the process of leptogenesis. Thus a complex Yukawa coupling is very much essential to yielding a non-zero lepton asymmetry. That necessity can be driven by the  CP phases that are supposed to be probed in either oscillation or neutrinoless double beta decay experiments. 

\begin{figure*}[t]
\begin{center}
\includegraphics[scale=0.37]{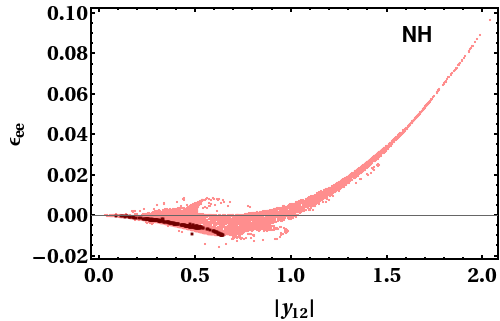}
\includegraphics[scale=0.36]{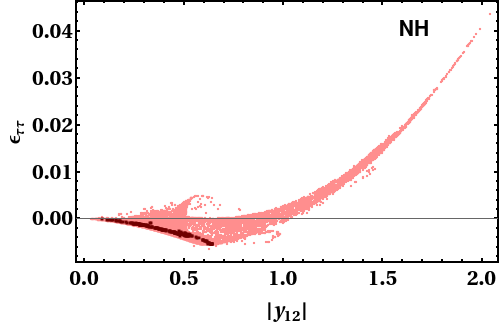} \\ 
\includegraphics[scale=0.37]{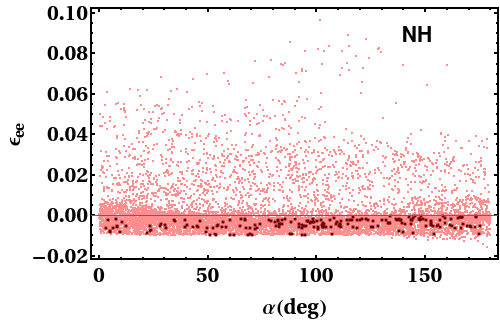}      
\includegraphics[scale=0.37]{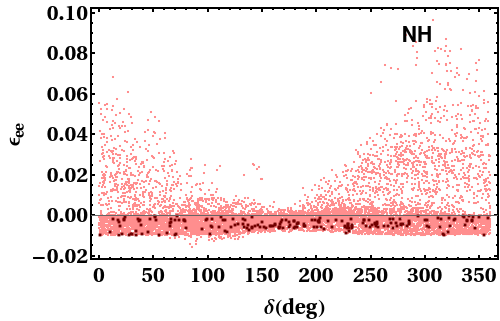}     
\caption{Lepton asymmetry as a function of the Yukawa couplings (upper panel) and low energy CP phases (lower panel) for NH. The dark-red points indicate the required values of the parameters needed to match the baryon to photon ratio $\eta_B = (6 - 6.18)\times 10^{-10}$.}
\label{fig:epsilonNH}
\end{center}
\end{figure*}

\begin{figure*}[t]
\begin{center}
\includegraphics[scale=0.38]{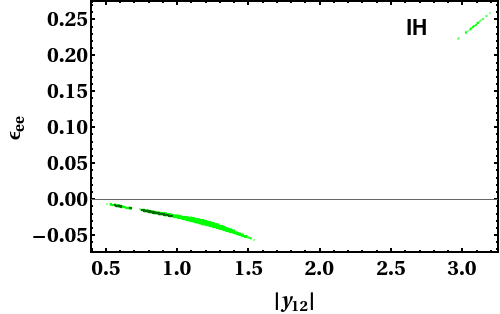}
\includegraphics[scale=0.37]{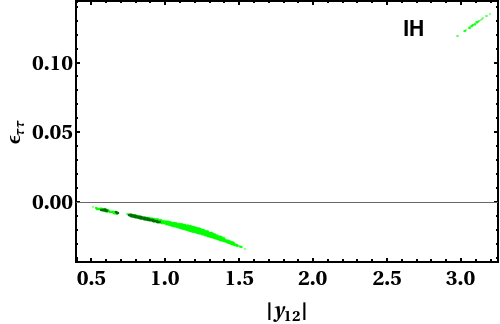} \\
\includegraphics[scale=0.38]{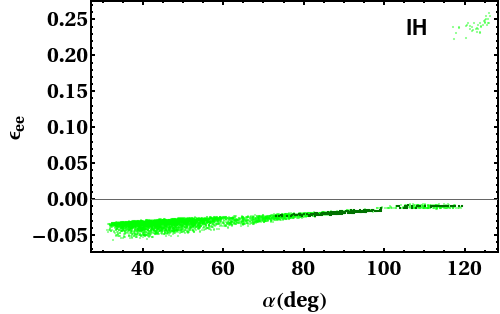}
\includegraphics[scale=0.38]{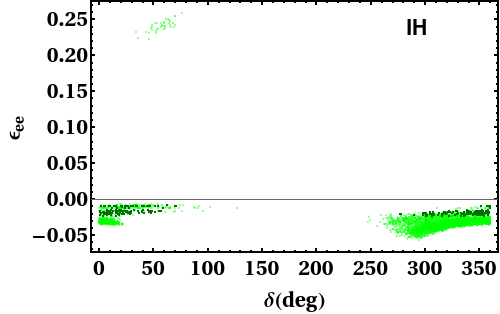} 
\caption{Leptonic CP asymmetries due to electron and tau flavor as a function of the Yukawa couplings (upper panel) and the Dirac and Majorana phases (lower panel) for IH. The dark-green points indicate the relevant data points satisfying the parameter space corresponding to $\eta_B = (6 - 6.18)\times 10^{-10}$.}
\label{fig:epsilonIH}
\end{center}
\end{figure*}

\begin{figure*}[t]
\begin{center}
\includegraphics[scale=0.25]{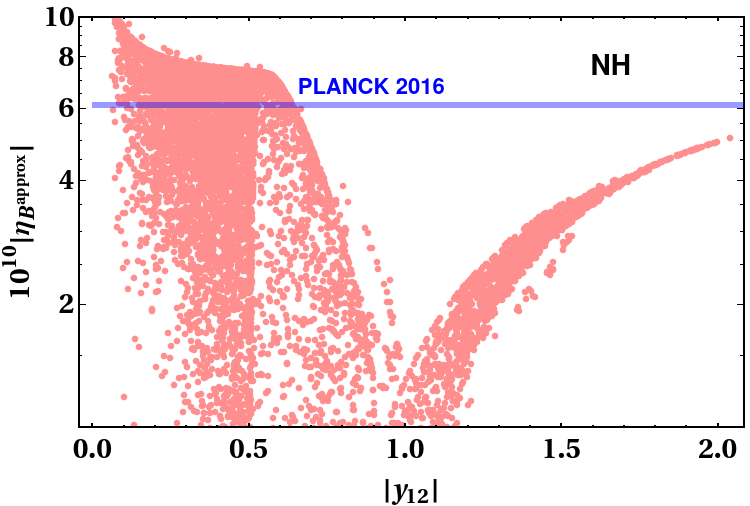}
\includegraphics[scale=0.25]{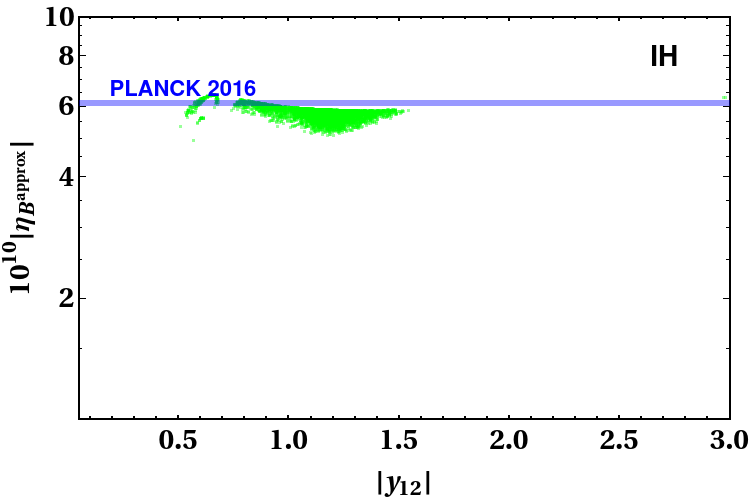}   
\caption{Baryon asymmetry as a function of Yukawa coupling ($y_{12}$) for NH (left panel) and IH (right panel). The violate line indicates the Planck bound reporting the baryon to photon ratio to be $\eta_B = (6 - 6.18)\times 10^{-10}$.}
\label{fig:etab}
\end{center}
\end{figure*}

In Figs.~\ref{fig:epsilonNH} and \ref{fig:epsilonIH} we present the variation of the lepton asymmetry associated with the electron and tau flavor with respect to the one of the Yukawa couplings ($y_{12}$). In the lower panel of the same, we show the dependency of the lepton asymmetry due to electron flavor as a function of the Dirac ($\delta$) and Majorana ($\alpha$) CP phases. Here, it is mentioned that the present analysis has very similar kind of predictions for all the lepton flavors in the context of lepton asymmetry when plotted as a function of the model parameters. 
Therefore, we keep the analysis open for discussing the variation of electron-flavored asymmetry with respect to the model parameters. As evident from Fig.~\ref{fig:epsilonNH} and Fig.~\ref{fig:epsilonIH} different hierarchies of neutrino masses reflect different kinds of correlations among the lepton asymmetry parameter and the neutrino Yukawa couplings.   
To have clarity on the order of the Dirac Yukawa coupling that satisfies both neutrino oscillation data and baryon asymmetry criteria, we have presented a numerical estimate of the same in Table \ref{tab:yukawa}, which has been constructed from Eq.~\eqref{eq:mD}. The corresponding values of the oscillation parameters have been mentioned in that table. It is to mention here that, the effective Yukawa couplings in the first two rows of the Dirac Yukawa matrix are suppressed by $\xi^4$ while those in the third row by $\xi^8$. This is also understandable from the matrix $\mathcal{D}$ of Eq.~\eqref{eq:mD}\footnote{While extracting the final result for baryon asymmetry we imposed perturbativity constraint on all the elements of the Yukawa coupling matrix (ensuring each $|Y_{\nu _{i \ell}}| \leq \sqrt{4\pi}$). Hence, the couplings that we show in the figures \ref{fig:epsilonNH},  \ref{fig:epsilonIH}, and \ref{fig:etab} fall within the perturbative limit.}.

\begin{figure*}
\begin{center}
\includegraphics[scale=0.35]{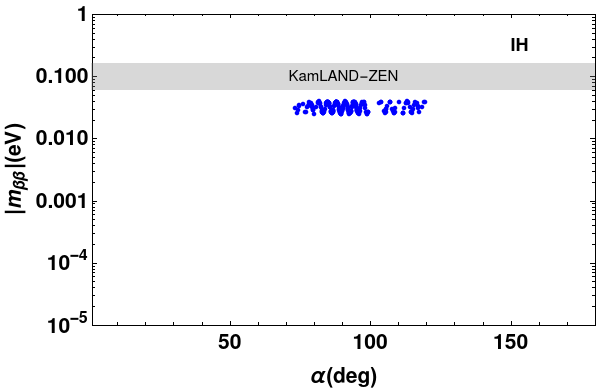}
\caption{Effective neutrino mass predicted by the model parameter space satisfying $\eta_B = (6 - 6.18)\times 10^{-10}$. The grey band indicates the limit on $m_{\beta\beta}$ set by the KamLAND-ZEN experiment \cite{KamLAND-Zen:2016pfg}, which falls in the range $61-165$ meV.}
\label{fig:ndbd}
\end{center}
\end{figure*}

 Since the baryon asymmetry criteria has imposed restriction on the CP phases of the PMNS matrix, it is instructive to study the effective neutrino mass parameter which influences the half-life of neutrino-less-double-beta decay (NDBD) amplitude \cite{Mohapatra:1998ye}. The effective neutrino mass governing the neutrinoless-double-beta decay~\cite{RevModPhys.59.671} is given by
\begin{equation}
m_{\beta\beta}= \sum_i \left| U_{ei}^2 m_i\right|,
\end{equation}

where, $i = 1,\,2,\,3$, $U$ being the lepton mixing matrix. In this framework, the effective mass would not receive any contribution from the lightest neutrino mass, as the corresponding mass eigenvalue is vanishing.  In Fig.~\ref{fig:ndbd} we have shown the variation of effective neutrino mass with the Majorana phase, the range of which is restricted by the baryon asymmetry criteria.  This figures evinces that the model prediction for the effective neutrino mass for the IH case lies within the bound as set by KamLAND-ZEN. Trivially, in our framework for NH also $m_{\beta\beta}$ falls well within the KamLAND-ZEN bound.  The IH case is still predictive as we can observe the constraint on the range of the Majorana phase, which is however not the case for NH. 
%%% new plot%%%%
%=====================================================
\begin{table}[h!]
\begin{scriptsize}
\begin{center}
\small
\begin{tabular}{|c | c | c |c |c |c | c | c | c | }
\hline
& $s_{13}^2$&$s_{12}^2$&$s_{23}^2$&$\Delta m_{31}^2/ 10^3$ &$\Delta m_{21}^2/ 10^5$&$\alpha$/$^0$&$\delta$/$^0$ & Yukawa Coupling ($Y_{\nu_{i\ell}}$)\\
\hline 
NH  &0.021 & 0.27 & 0.46 & 2.58 eV$^2$ & 7.66 eV$^2$ & 113 & 141.5&           
$ \small{ \left(
\begin{array}{ccc}
0.7+ 0.766i& -0.59 - 0.37i& -0.59 - 0.37i\\
0.15 - 0.36i& -0.04 + 0.42i& -0.04+ 0.42i\\
 -i & i & i\\
\end{array}
\right)}$\\
\hline
%\hline
IH & 0.022& 0.34 & 0.45 & 2.4 eV$^2$ & 7.14 eV$^2$& 115 & 15.87 & $\left(
\begin{array}{ccc}
-0.594 - 0.018 i& 0.59 + 0.019 i & 0.73 - 0.3 i\\
 -1.28 + 2.42i & 1.28-2.42 i&1.41-3.08 i\\
 -i & i & i \\
\end{array}
\right)$ \\
\hline
\end{tabular}
\caption{ Dirac Yukawa coupling matrices for NH (top panel) and IH (bottom panel), satisfying both neutrino oscillation data and the PLANCK bound on baryon to photon ratio. Yukawa couplings have been generated using the input parameters mentioned in the table itself. These sets of neutrino oscillation parameters and the corresponding Yukawa matrices yield the baryon to photon ratio to be, $\eta_B\,=\,6.04 \times 10^{-10}$(for NH) and $\eta_B\,=\,6.15 \times 10^{-10}$(for IH) respectively.}
\label{tab:yukawa}
\end{center}
\end{scriptsize}
\end{table}
%=============================================================================================

We mention here that, the different flavored asymmetry parameters along with the washout parameters behave similarly to the low-energy CP phases. This is due to the reason that both of these parameters depend mainly on the complex Yukawa couplings, the complex nature of which is solely because of the existence of the low-energy CP phases. To have a clear vision of the role of the individual CP phases on the non-zero lepton asymmetry, we identify the allowed regions of the Dirac and Majorana phases which altogether yield a  correct sign for the CP asymmetry and account for a sufficient amount of baryon asymmetry. Those identified regions for these phases are presented in dark color which can be found in the lower panels of the Figs. \ref{fig:epsilonNH} and \ref{fig:epsilonIH}. In order to identify these regions we impose restrictions on the complete parameter space (consists of the neutrino mixing angles and the CP phases) with the condition that one should get the CP asymmetry with a correct sign and the constrained region should also give rise to the desired order of the baryon asymmetry. For NH, we do not see any restrictions on either $\delta$ or $\alpha$. However, for IH one can notice in the lower panel of Fig.~\ref{fig:epsilonIH} that, $\delta$ and $\alpha$ are having  restricted ranges.  One can notice two allowed regions of $\delta$ from the entire range of $0\,-\,360^{\circ}$ one from $0.2^{\circ}\,-\,69^{\circ}$ and another from $276^{\circ}\,-\,359^{\circ}$. At the same time $\alpha$ is allowed to oscillate in the ranges $73^{\circ}\,-\,99^{\circ}$ and $102^{\circ}\,-\,119^{\circ}$ from the aforementioned range which has been used as numerical input. 
%This leads us to comment that the case with normal hierarchy is more predictive in the context of the restriction imposed by the baryon asymmetry data. Moreover, it is quite anticipating to note that this $\eta_B$ criteria also put restrictions on the $3\sigma$ range of the atmospheric mixing angle ($\theta_{23}$). 

%We notice a preference for the higher octant of $\theta_{23} \,> \pi/4$. The constraint over the $\delta\,-\,\theta_{23}$ and $\alpha\,-\,\theta_{23}$ planes are demonstrated in fig.~\ref{fig:s23}.}

We present the model prediction for the baryon-to-photon ratio as a function of a Yukawa parameter in Fig.~\ref{fig:etab} using the first approach as mentioned above. This figure illustrates that for both the mass hierarchies of light neutrinos, adopting the analytically approximated solution from the BEQs we get a wider region of model parameter space which satisfies the $\eta_B$ constraint. 
%\clearpage

%===============================================
\subsection{Boltzmann equations and solutions} \label{sec:beq}
% NEW EQUATIONS

To determine the lepton asymmetry in the present epoch, one has to solve the BEQs for lepton number density which in turn depends on the instantaneous value of the RH neutrino number density. As mentioned before, here we have chosen the RHN masses to be nearly degenerate, and within a few TeV. Thus in such a mass window, it is instructive to  study resonant leptogenesis from the decay of the RHNs. This temperature regime at the same time also allows looking for the lepton flavor effects, as the individual lepton flavors become distinguishable.  In such a scenario one has to take the flavor effects into account while solving the coupled BEQs also. Thus the required set of classical kinetic equations taken from the Ref.~\cite{Pilaftsis:2005rv} governing the evolution of RH neutrino number density and the lepton flavor density are given by,
%\begin{small}
\begin{eqnarray}
\label{beq1} 
\frac{d \eta_{N_{i}}}{dz} &=& \frac{z}{H(z=1)}\
\Bigg[\,\Bigg( 1 \: -\: \frac{\eta_{N_{i}}}{\eta^{eq}_{N_{i}}}\,
\Bigg)\, \sum_{k\,=\,e,\mu,\tau} \bigg(\,
\Gamma^{D\; (i k)} \: +\: \Gamma^{S\; (i k)}_{Yuk}\: +\:
\Gamma^{S\; (i k)}_{Gauge}\, \bigg) \nonumber\\ &&-\,
\frac{2}{3}\, \sum_{k\,=\,e,\mu,\tau} \eta^{k}_{\ell}\,
\varepsilon^{k}_{i}\, \bigg(\,
\widehat{\Gamma}^{D\; (i k)} \: +\: 
\widehat{\Gamma}^{S\; (i k)}_{Yuk} \: +\:
\widehat{\Gamma}^{S\; (i k)}_{Gauge}\, \bigg)\,\Bigg]\,,\\[5mm]
  \label{beq2} 
\frac{d \eta_\ell}{dz} &=& 
\frac{z}{H(z=1)}\, \Bigg\{\, \sum\limits_{i=1}^2\,
\varepsilon^{\ell}_{i}\ \Bigg(
\frac{\eta_{N_{i}}}{\eta^{eq}_{N_{i}}} \: -\: 1\,\Bigg)\, 
\sum_{\beta\,=\,e,\mu,\tau} \bigg(\,
\Gamma^{D\; (i k)} \: +\: \Gamma^{S\; (i k)}_{Yuk}\:
+\: \Gamma^{S\; (i k)}_{Gauge}\, \bigg) \nonumber\\
&&-\,\frac{2}{3}\, \eta_\ell, \Bigg[\, \sum\limits_{i=1}^2\,
B^{\ell}_{i}\,
\bigg(\, \widetilde{\Gamma}^{D\; (i \ell)} \: +\: 
\widetilde{\Gamma}^{S\;(i \ell)}_{Yuk}\: +\: 
\widetilde{\Gamma}^{S\; (i \ell)}_{Gauge}\: +\: 
\Gamma^{W\; (i \ell)}_{Yuk} + \Gamma^{W\;(i \ell)}_{Gauge}
\,\bigg)\nonumber\\
&& \qquad\qquad\,\,\;
\: +\: \sum_{k\,=\,e,\mu,\tau}\,\bigg(\,\Gamma^{\Delta L =2\:(\ell
k)}_{Yuk} \: +\:\Gamma^{\Delta L =0\:(\ell k)}_{Yuk}\,\bigg) \Bigg] \nonumber\\
&&-\, \frac{2}{3}\,\sum_{k\,=\,e,\mu,\tau} \eta_{k}\,
\Bigg[\,\sum\limits_{i=1}^2\,k_{i}
\bigg(\,\Gamma^{W\;(i k)}_{Yuk}\:
 +\: \Gamma^{W\; (i k)}_{Gauge}\,\bigg)\nonumber\\ 
 && \qquad\qquad\qquad\qquad\qquad\quad\,\, \: +\: \Gamma^{\,\Delta L =2\:
(k \ell)}_{Yuk} \: -\: \Gamma^{\,\Delta L =0\: (k \ell)}_{Yuk}
\Bigg]\,\Bigg\}\,,
\end{eqnarray}
%\end{small}

% \begin{gather}\label{eq:beq}
%  \frac{d\delta\eta_{N_i}}{dz} = \frac{\mathcal{K}_1(z)}{\mathcal{K}_2(z)}\left[1+ (1- K_i z)\delta\eta_{N_i}\right] ~~~ \text{with}~~~ i =2,3\\ 
%  \frac{d\eta_{l}}{dz} = z^3 \mathcal{K}_1(z) K_i\left( \delta \eta_{N_i} \epsilon_{il} -\frac{2}{3} B_{il}\eta_{l}\right), ~~~ \text {with}~~~ l = e,\mu,\tau,
% \end{gather}
%$ K_i = \frac{\Gamma_{N_i}}{\zeta(3)H_N}$ as the washout parameter associated with the i'th RHN of the model and 

 In the above equations, the several $\Gamma$ functions indicate the reaction densities that correspond to the various washout processes. A detailed description of this can be found in Appendix \ref{sec:Appendix B}.
Here, $i = 1,2$ implies the involvement of the heavier RHN states ({\it i.e.}, $N_1$ and $N_2$ here) in producing the discussed lepton asymmetry and $\ell,\, k$ represent the lepton flavors\footnote{from here onwards we identify the Dirac Yukawa coupling $Y_{\nu_{i}}^\ell$ simply as $y_{i \ell}$.}. In order to have an idea of the deviation of RHN abundance from the equilibrium abundance one can define $\delta \eta_{N} = \frac{N_{N}}{N^{\rm eq}_N} -1$ and substitute in the Eqs. \eqref{beq1} and \eqref{beq2}. This results in the evolution of the deviation from the equilibrium abundance of the RHNs with respect to the temperature function $z ( = M_N /T)$. In the above $B_{i}^\ell = \frac{|Y_{\nu_{i}}^\ell|^2}{(Y_{\nu}Y_{\nu}^{\dagger})_{ii}}$. Here $N_N^{\rm eq}$ denotes the equilibrium number density of the RHNs under discussion. It is to note here that, this equilibrium abundance is normalized to the photon number density and one can write,
$$\eta_N^{\rm eq} \approx \frac{1}{2}z^2 \mathcal{K}_2(z).$$
The photon number density is parametrized as $\frac{2M_N^3}{\pi^2} \frac{\zeta(3)}{z^3}$ where, $\zeta(3)$ is the Apery's constant which is nearly equal to 1.202.
  The temperature dependence of the above BEQs is expressed by means of the dimensionless parameter $z = M_N /T$.
The above set of coupled BEQs (Eq. \eqref{beq1} and \eqref{beq2}) can be numerically solved up to a value of z where the quantity $\eta_\ell$ attains a constant value. On solving the above BEQs, the resulting lepton charge abundance can be transferred to the baryon density using the following equation.
\begin{equation}\label{eq:etaL}
\eta_B = -\frac{28}{51} \frac{1}{27}\sum_{\ell} \eta_\ell,
\end{equation}
with, $\eta_\ell$ being the yields for $\ell$th flavored lepton charge density obtained from the solution of the above BEQs. The factor of 28/51 emerges from the fraction of lepton asymmetry reprocessed into a baryon asymmetry by the electroweak sphalerons while 1/27 is the dilution factor from photon production until the recombination epoch.

\begin{table}[h!]
\begin{center}
\begin{tabular}{| c | c | c | c | c | c | c |c|  }
  \hline
BP  &$\epsilon_e$ & $\epsilon_\mu$ & $\epsilon_\tau$&  $ K_1$ & $ \eta_B =  -\frac{28}{51} \frac{1}{27}\sum_{\ell} \eta_\ell $   \\
\hline 
I &   $0.0045$  & $-0.0045$   & $-0.0026$& $2.38 \times 10^5$ & $2.13 \times 10^{-10}$ \\
  \hline
\end{tabular}
\caption{Benchmark values for the lepton asymmetry parameters and the corresponding washout amount which altogether yield the final baryon to photon ratio ($\eta_B)$. This $\eta_B$ value is evaluated at $z = 20$ from the solution of the BEQs.}
\label{tab:etab}
\end{center} 
\end{table}
\begin{figure*}[t]
\begin{center}
\includegraphics[scale=0.45]{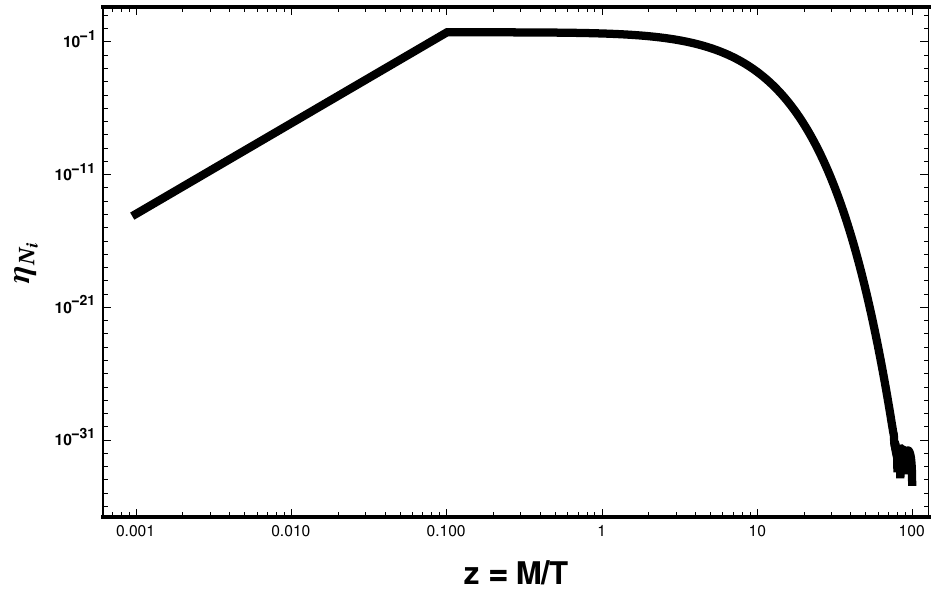}
\includegraphics[scale=0.25]{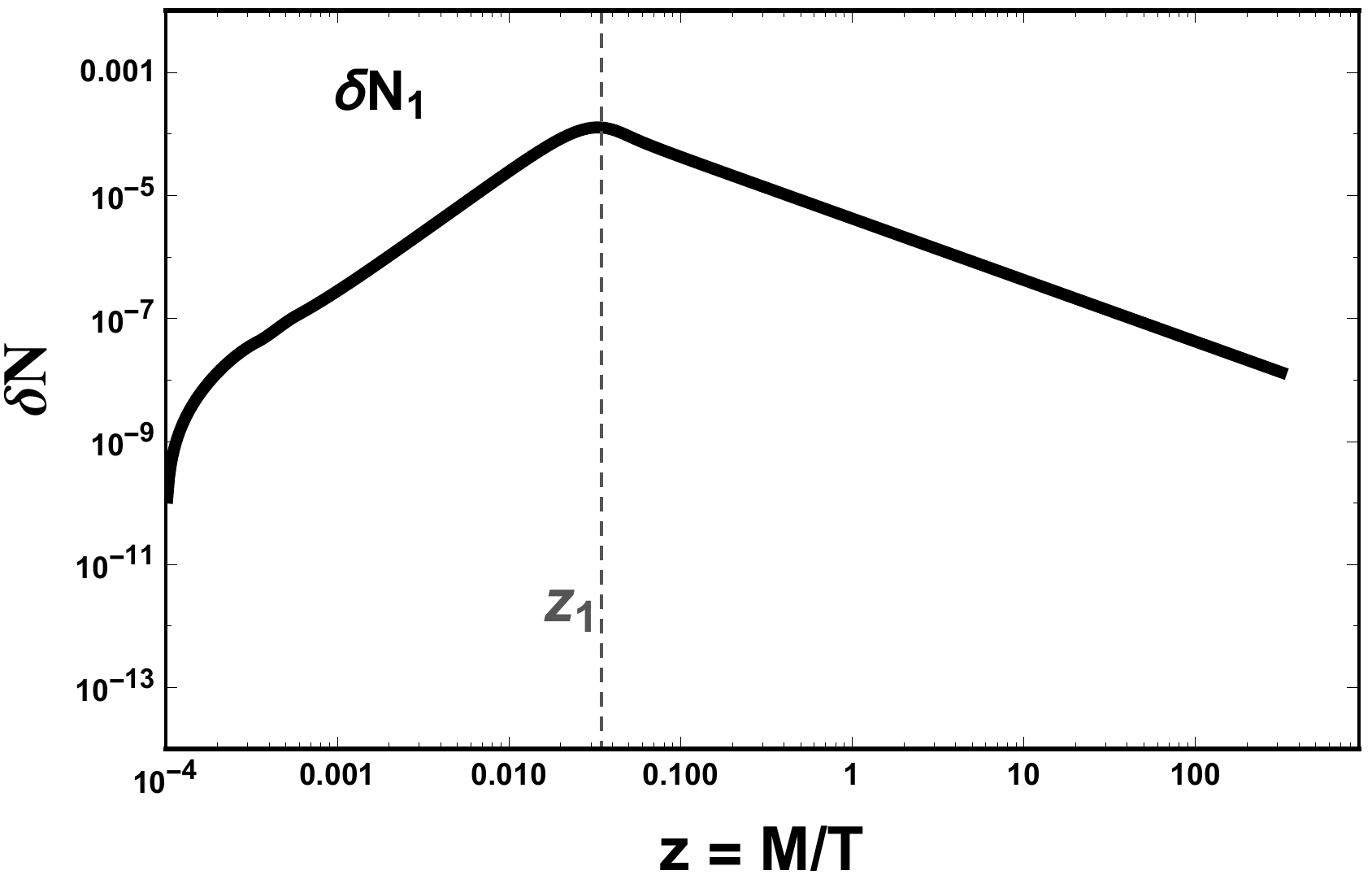}
\includegraphics[scale=0.25]{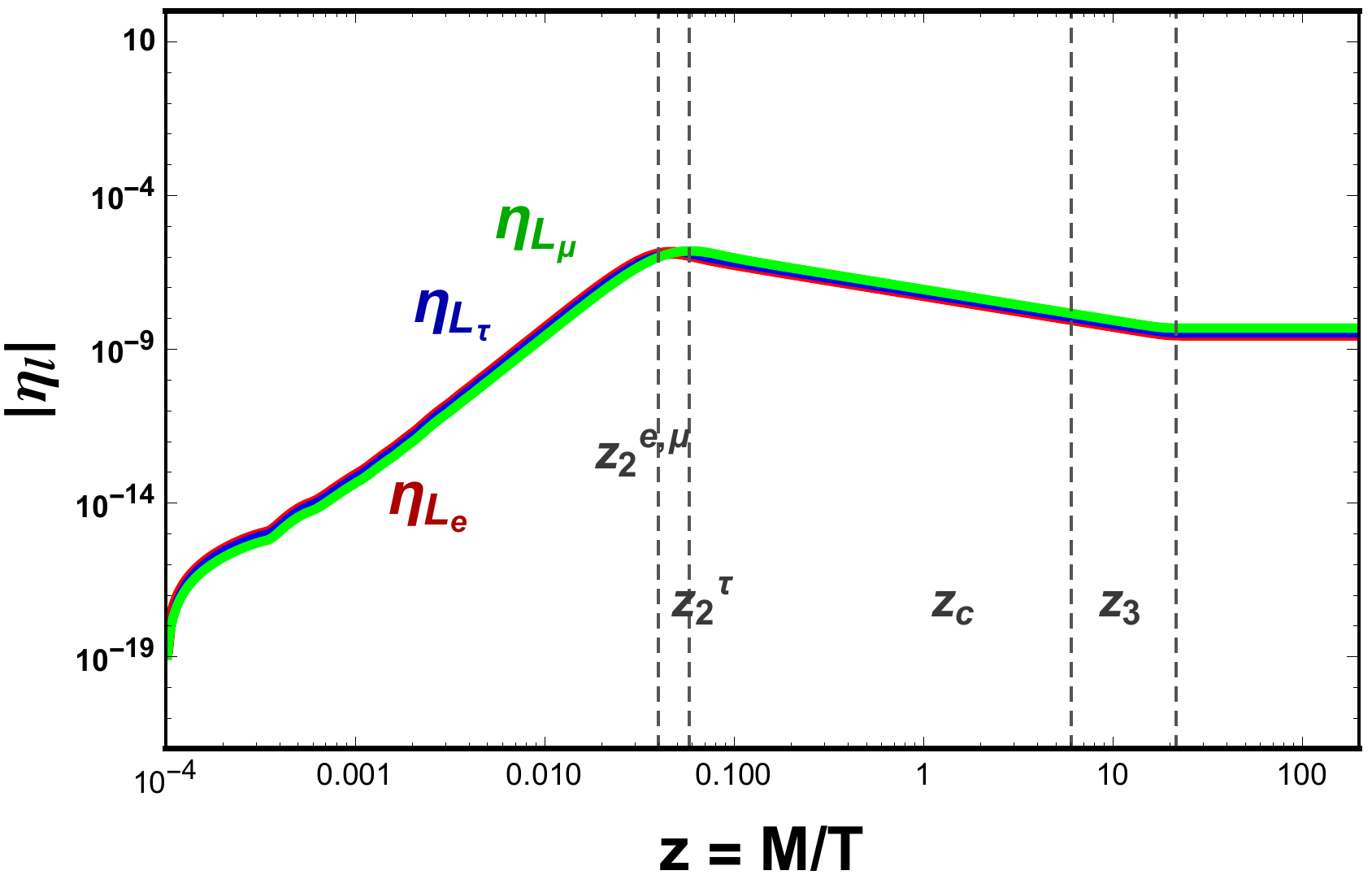}   
\caption{In the upper panel evolution of the RHN abundance (left) and deviation from the equilibrium RHN abundance (right) are shown. In the bottom panel, we show the evolution of the lepton charge densities through varying $z$. This figure is generated using the benchmark values tabulated in Table \ref{tab:etab}. The washout amount associated with all the lepton flavors essentially satisfies $ \kappa_\ell^{\rm eff} \geq 5$, being $ \kappa_{e,\,\mu,\,\tau}^{\rm eff} \approx10^5$.}
\label{fig:NBL}
\end{center}
\end{figure*}
%These two initial conditions together establishes the fact that any asymmetry generated while thermally populating the RHN is quickly washed out once a thermal population is reached. 
 In order to solve the above set of BEQs one can usually choose either of two different sets of initial conditions, i) $\eta_{N} = 0$ and $\eta_{\ell} = 0$, also familiar as vanishing initial abundance (VIA) and ii)  $\eta_{N} = \eta_N^{\rm eq}$ and $\eta_{\ell} = 0$, known as thermal initial abundance (TIA) \cite{Granelli:2020ysj}. The first set ensures zero initial abundance of the RHNs whereas the later set implies the maximum/equilibrium abundance of the RHNs. The second condition on $\eta_{\ell}$ ensures the non-existence of any primordial lepton asymmetry. Choosing the VIA condition is more suitable to the weak washout scenario, due to the fact that in absence of equilibrium abundance of the RHNs, the production of the same is assumed to take place via the $\Delta L\,=\,1$ processes \cite{Buchmuller:2004nz}\footnote{Despite the fact that we have encountered a washout regime which is strong enough (as $K \approx \, 10^5$) we have considered the VIA condition for solving the set of BEQs for completeness. It is well known for a strong washout case that, the final asymmetry would remain unaltered for either choice of initial conditions (see \cite{Dolan:2018qpy} for comparison on the choice of initial condition). From that perspective, the choice of TIA of the RHNs would be more natural in our analysis led by the characteristic feature of this HDO-motivated model.}.  The VIA condition implies the RHN production by the $\Delta L\,=\,1$ processes which include both Yukawa and gauge boson mediated $2\leftrightarrow 2$ scatterings.

 The final lepton asymmetry may get affected by the washout produced by the aforementioned
$2 \leftrightarrow 2$ scatterings which violate the lepton number by $\Delta L \,=\,0,\,1,\,2$, provided their strengths are larger than that of the inverse decay. Among them, the $\Delta L\,=\,1$ process mainly enhances the RHN production and also the efficiency factor. From  Ref. \cite{Buchmuller:2004nz} it is clear that the Yukawa and gauge boson mediated scatterings are important only for the scenarios of weak washout which are realized by the condition $K( = \Gamma_i / H_N) < 1$, K being the washout factor. 
This is due to the reason that, in such
a scenario of weak washout the final asymmetry is strongly dependent on the initial condition related to
the RHN production.

In our analysis, we encounter a very strong washout regime where $K$ comes out to be around $10^5$ or higher. It is also evident from \cite{Buchmuller:2004nz} that for $K >> 1$ the final asymmetry is independent of the initial condition. For such a huge washout as in our case, inverse decays play the most dominant role among
all the washout processes including the $\Delta L= 0,\,1,\,2$ scatterings. Moreover, the cross sections of $\Delta L= 0,\,2$ scatterings are suppressed by the fourth power of the Yukawa couplings and thus smaller in order of magnitude (please refer to the appendix \ref{sec:Appendix B}) from the decay or inverse decay (ID) width, $\Gamma_D$\footnote{IDs being the most dominant among all the washout processes it is sufficient to consider only the IDs and hence the inclusion of scatterings would leave insignificant changes to the final lepton asymmetry \cite{Buchmuller:2004nz}.}. For clarity and a better understanding of this fact, we tabulate the numerical estimates of the reaction densities of all the $2 \leftrightarrow 2$ processes at $z\,=\,1$ and $z\,=\,10$ in Table \ref{tab:washout}. The comparison among several reactions densities corresponding to each washout process gives us the notion that after a certain period of time after $z\,=\,1$ the scatterings become weaker in strength than the inverse decay\footnote{We have also checked that our prediction in this regard matched with the results provided in Refs. \cite{Pilaftsis:2005rv,Granelli:2020ysj}.}. This further allows us to believe that the IDs play the most decisive role in a strong washout scenario in the evolution of final lepton asymmetry from which the baryon asymmetry at the present day is calculated. 
\begin{table}[h!]
\begin{center}
\begin{tabular}{| c | c | c |  }
\hline
Reaction densities ($\Gamma _X^Y$) & $z = 1$ & $z = 10$\\
\hline
ID $(\Gamma ^N_{\ell \phi})$& 1.44$\times 10^{-7}$& $4.4 \times 10^{-8}$\\
  \hline 
$\Delta L\,=\,0$ ($ \Gamma^{\,\Delta L =0\:
(k \ell)}_{Yuk}$) &  $ 1.8 \times 10^{-8}$& $ 5.1 \times 10^{-11}$\\
\hline 
  $\Delta L\,=\,1$ ($\Gamma^{S\; (i k)}_{Yuk} \,\,\text{ and}\,\,\Gamma^{S\; (i k)}_{\rm Gauge}$)  & $3.5 \times 10^{-8}$& $5.5 \times 10^{-11}$\\
  \hline
   $\Delta L\,=\,2 $ ($ \Gamma^{\,\Delta L =2\:
(k \ell)}_{Yuk}$)  & $9.1 \times 10^{-18}$& $2.7\times 10^{-45}$\\
   \hline
\end{tabular}
\caption{The above reaction densities have been evaluated using the benchmark point which yields the lepton and baryon asymmetry of the desired order as mentioned in Table~\ref{tab:etab}.  For calculating the reaction densities we have used the relevant expressions from Appendix~\ref{sec:Appendix B}.}
\label{tab:washout}
\end{center} 
\end{table}

In Fig.~\ref{fig:NBL} we show the evolution of RHN number density and lepton flavor density, obtained from the solution of the BEQs. This figure tells us that, after a certain temperature the $\eta_\ell$ number densities due to individual lepton flavors attain a saturation. In the upper left panel of this figure, we show the evolution of the RHN abundance starting from zero initial abundance ($ \mathcal{O} (10^{-15}$)) and in the right, we show the evolution of the deviation from the equilibrium RHN abundance. It is evident from the left figure that  the RHN abundance gradually reaches the thermal or the equilibrium abundance at around $z \,=\,0.1$.
One can notice from the lower panel of Fig.~\ref{fig:NBL}, that the numerical solution for $\eta_\ell$ through the BEQ
exhibits different behavior in the three kinematic regimes, characterized by the specific values of the parameter $ z = m_N / T$. Based on this fact there are three benchmark values of $z$ signifying three temperature regimes. Analytically these temperature regimes can be approximated by writing them as a function of the washout factors associated with each lepton flavor as 
$$ z_2^\ell \approx 2(K_\ell^{\rm eff})^ {-1/3}, \;\;\; z_3^\ell \approx 1.25 \text{ln} ( 25 K_\ell^{\rm eff}), $$
which are found to be true for the set benchmark values of the respective parameters mentioned in Table \ref{tab:etab}.
A noticeable feature of this model is, for $e$ and $\mu$ flavors the lepton charge densities evolve exactly in a similar fashion, which however slightly differs for the tau flavor. Also, it is to note here that the final lepton asymmetry receives almost an equal contribution from the Yukawa couplings associated with each individual lepton flavor.

%===========================================================

\section{Collider prospects: RHN pair production}
\label{sec:collider}

In the previous sections, we mostly discuss how our model can generate the observed 
BAU through the resonant leptogenesis mechanism. This has been achieved by considering two almost degenerate TeV-scale RHNs (denoted by $N_1$ and $N_2$) in the spectrum of our model. There is another light keV-scale neutrino $N_3$ (predominantly right-handed in nature) present in the model but it has no role to play in the leptogenesis mechanism. In our previous paper~\cite{Choudhury:2020cpm}, we discussed how $N_3$ can act as a dark matter candidate and briefly discussed the collider prospects of $N_{1,2}$. Here, we study the collider signatures of the TeV-scale $N_{1,2}$ in more detail. 
Usually, the parameters in the neutrino sector are not very sensitive to collider experiments.
The light-heavy mixing angle is one parameter that can contribute to BAU as well as can control the production rates of RHNs at colliders.
These RHNs are not charged under the SM gauge group and 
therefore, direct productions of them through their interactions with the SM fields are very suppressed (can only arise through their mixing with the SM neutrinos~\cite{Deppisch:2015qwa,Banerjee:2015gca}). 
We obtain the light-heavy mixing element, e.g. 
$|V_{\mu N_{1,2}}|^2\sim 10^{-4}-10^{-2}$ for the parameter space corresponding to observed BAU. This range of $|V_{\mu N_{1,2}}|^2$ can be accessible at the future muon collider experiments (see the recent paper \cite{Li:2023tbx} and references therein). Various signatures of heavy RHNs that depend on the light-heavy mixing in the context of the future lepton colliders can be found in Refs.~\cite{Basso:2009hf,Li:2023tbx,Banerjee:2015gca,Chakraborty:2018khw} and in the context of the LHC experiments in 
Refs.~\cite{delAguila:2007qnc,Alva:2014gxa,Pascoli:2018heg}.

Apart from the direct production of RHNs controlled by the light-heavy mixing, they can also be produced through the decay of another BSM particle~\cite{ThomasArun:2021rwf,Arun:2022ecj,Bhaskar:2023xkm}.
In our model, $N_{1,2}$ can be pair produced in through the decay of the TeV-scale $\mathrm{U}(1)$ gauge boson $Z'$ leading to a new production mode for RHNs. This channel depends mostly on the parameters related to the gauge sector of the model that has very little role to play in neutrino phenomenology. In the subsequent discussion, we investigate interesting signatures of our model that can be tested at the LHC experiments.

\begin{figure}[h!]
\begin{center}
\includegraphics[scale=.7]{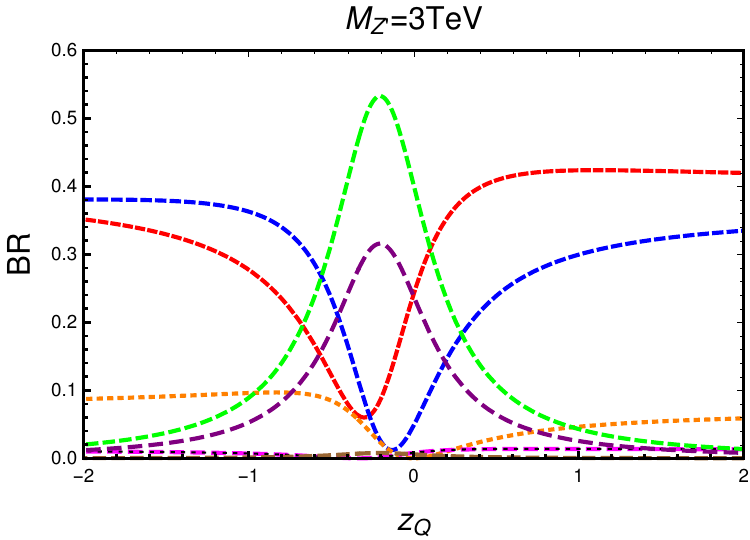}\\
\includegraphics[scale=1]{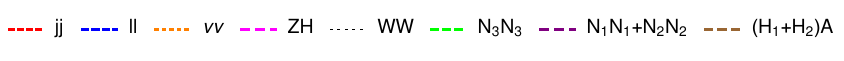}
\caption{Branching fractions of possible two-body decay modes of $Z'$ as functions of the free charge $z_Q$. For these plots, we choose the benchmark parameters as follows: $g_z=0.3$, $M_{N_1}=M_{N_2}=1$ TeV, $M_{H_1}=M_{H_2}= 1$~TeV. Here, $j$ includes all quarks except top-quark and $\ell$ includes all charged leptons.}
\label{Branching_3TeV}
\end{center}
\end{figure}

For the LHC studies, the relevant free parameters of our model are $M_{Z'}$ (mass of the $Z'$ boson), $M_{N_{1,2}}$ (masses of $N_1$ and $N_2$ RHNs. Since they are almost degenerate, leads to one free mass), $\mathrm{U}(1)_z$ gauge coupling $g_z$ and free charge $z_Q$. Other parameters like VEVs $x_i$ are set to some benchmark values. Heavy scalar masses $M_{H_{1,2}}$, mixing angle in the scalar sector etc. have a very little role in collider signatures that we intend to discuss here. The BRs of $Z'$ to various modes are presented as functions of $z_Q$ in Fig.~\ref{Branching_3TeV}. To obtain this plot, we assume $M_{Z'}=3$ TeV, $g_z=0.3$, $M_{N_{1,2}}=1$ TeV and $M_{H_{1,2}}=1$~TeV. Note that the $Z'$ BRs are insensitive to the choice of $g_z$ and also to $M_{Z'}$ (as long as the decay to BSM particles is kinematically open). As we mentioned above, the possibility of observing the pair production of RHNs through $Z'$ decay will open up if the $Z'\to NN$ BR is large enough such that it is within the reach of the HL-LHC. In Fig.~\ref{Branching_3TeV}, we see that around $z_Q\sim -1/3$, $Z'\to N_3N_3$ BR becomes 50\% and $Z'\to N_1N_1+N_2N_2$ BR shares almost 30\%. As a consequence, the other standard decay modes of $Z'$ namely dijet, dilepton etc. will be suppressed around $z_Q\sim -1/3$ value. Although the $Z'\to N_3N_3$ mode has the highest BR around $z_Q\sim -1/3$, this mode has relatively less prospect to be observed at the LHC.  Since $N_3$ is very light and it has very weak couplings to the lighter states, they essentially decay outside the detector. Therefore, the process 
$pp\to Z'\to N_3N_3$ will lead to a missing energy signature. This particular signature can be observed if there is any accompanying visible particle like a jet. Other electroweak gauge bosons like a photon or a $W$ or a $Z$ boson can also appear with the missing energy. However, they are suppressed compared to the mono-jet plus missing energy mode.

\begin{figure*}[]
\begin{center}
\captionsetup[subfigure]{labelformat=empty}
\subfloat[\quad\quad(a)]{\includegraphics[scale=0.55]{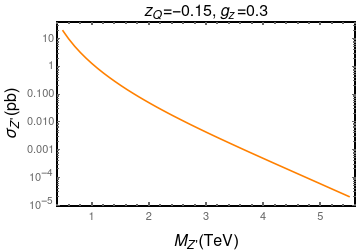}\label{fig:cxmzp}}
\subfloat[\quad\quad(b)]{\includegraphics[scale=0.55]{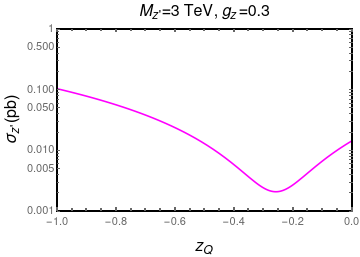}\label{fig:cszq}}\\
\subfloat[\quad\quad(c)]{\includegraphics[scale=0.55]{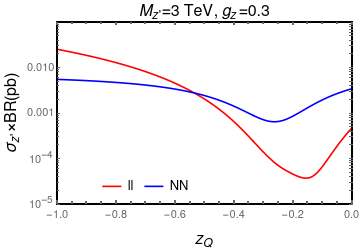} \label{fig:csdilep}}
\caption{Production cross section for $Z'$ as functions of (a) $M_{Z'}$ with fixed $z_Q$ and (b) $z_Q$ with fixed $M_{Z'}$. (c) Production cross section of $Z'$ times dilepton BR as functions of $z_Q$. All plots are made by choosing $g_z=0.3$.}
\label{fig:prodcs}
\end{center}
\end{figure*}

In Figs.~\ref{fig:cxmzp} and \ref{fig:cszq}, we show the $Z'$ production cross section 
$\sg_{pp\to Z'}$ at the 14 TeV LHC as functions of $M_{Z'}$ and $z_Q$ respectively. The couplings of $Z'$ with the SM quarks depend on the free charge $z_Q$ and we see that around $z_Q\sim -0.25$, the $\sg_{pp\to Z'}$ acquires the minimum value. Actually, the behavior of $\sg_{pp\to Z'}\times \textrm{BR}$ for dilepton and di-RHN modes as functions of $z_Q$ is more important to check. Since the dilepton resonance search data provides the most stringent bounds, we need to minimize that quantity to satisfy the constraints. At the same time, we also want the di-RHN mode to be sufficiently large to
observe it in collider experiments. 
This happens around $z_Q\sim -0.15$ as can be seen in~\ref{fig:csdilep}.  
Later, we will see that the region $z_Q\sim -0.15$ is the most favored region which can be probed through the same-sign di-RHN channel. Note that this channel is a unique channel that can be used to probe the simultaneous presence of $Z'$ and RHN. However, if we talk about the discovery prospect of $Z'$ alone, the dilepton and dijet decay modes of $Z'$ obviously perform better than the di-RHN mode i.e. $Z'$ would possibly be discovered in the dilepton/dijet modes before the di-RHN mode. In leptophobic $\mathrm{U}(1)$ context, in Ref.~\cite{Arun:2022ecj} it was shown that  a considerable region of the parameter space beyond the reach of the future dijet searches can be discovered exclusively using the di-RHN channel. Comparing the dijet with the dilepton channel, for a sequential $Z'$, the $\sg\times BR$ exclusion in the dilepton channel~\cite{Aad:2019fac,Sirunyan:2021khd} is roughly one order smaller than the dijet channel~\cite{Aad:2019hjw,Sirunyan:2019vgj}.

\subsection{Leptonic final states}

One of the most striking features of the model, as discussed in~\cite{Choudhury:2020cpm}, is that there is a large coupling of the $Z'$ to the RHNs due to their $\mathrm{U}(1)_z$ charges. Although $N_1$ and $N_2$ are significantly heavier compared to $N_3$, the branching fractions of these heavy RHNs are seen to be competitive with the lighter ones as shown in 
Fig.~\ref{Branching_3TeV}. Furthermore, the RHNs $N_{1,2}$ can decay into the following SM states through their mixing with the SM neutrinos: $N_{1,2}\to W^{\pm}\ell^{\mp}, ~Z(\nu+\bar{\nu}),~H(\nu+\bar{\nu})$ (where $H$ is the SM Higgs boson and $H_{1,2}$ are taken to be heavier than $N_{1,2}$ such that the corresponding decay is kinematically forbidden).
This opens up many interesting signatures that can potentially be observed in future LHC runs. Below, we categorize some of the channels in terms of the number of leptons ($\ell = e,\mu$) present in the final states.
\begin{itemize}
\item \textbf{Mono-lepton:} The mono-lepton signature can arise from the following decay
\begin{eqnarray}
pp \to Z'\to N_1N_1+N_2N_2 \left\{\begin{array}{ccl}
  &\to & (W^\pm_h\ell^\mp)(Z_h\nu) \\
 &\to & (W^\pm_h\ell^\mp)(H_h\nu)
\end{array}\right\}\ ,
\end{eqnarray}
where $W_h,Z_h,H_h$ denote the hadronic decays of those particles. These channels will essentially lead to one lepton, a couple of jets, and some amount of missing energy in the final state. The LHC prospect study of this channel is not performed in any phenomenological analysis before to the best of our knowledge. This channel is not fully reconstructible due to the presence of missing energy. However, one RHN can be reconstructed. In terms of BR, this signal will gain over the di-lepton and tri-lepton channels discussed below.  The SM background for this channel will be large and therefore very difficult to isolate this signal over the background. One interesting feature of this channel is that it contains two boosted fatjets arising from the hadronic decays of the massive SM bosons.
Therefore, one can make use of the jet-substructure technique to separate out this signal from the large background.

\item \textbf{Di-lepton:} The di-lepton signature can arise from the following decay

\begin{equation}
pp \to Z'\to N_1N_1+N_2N_2 \to (W^\pm_h\ell^\mp)(W^\pm_h\ell^\mp)\ .
\end{equation}
Note that a di-lepton final state can also arise from the decay of RHNs to 
$(Z_\ell\nu)(Z_h\nu)+(Z_\ell\nu)(H_h\nu)$ mode. First of all, this channel will contribute little due to the small leptonic branching of $Z$ and also not fully reconstructible. But a huge dilepton background arises from $pp\to Z\to \ell\ell$ and to suppress it a $Z$-veto cut is mandatory. This $Z$-veto cut will remove the signal containing leptonic decay of $Z$. Since the RHNs are Majorana in nature, they can violate the lepton number by two units and produce a smoking-gun same-sign dilepton signature~\cite{Das:2017flq,Das:2017deo}. Therefore, the di-lepton channel can further be categorized into same-sign and opposite-sign dilepton final states. The background for the opposite-sign dilepton channel is large but the same-sign dilepton channel is almost background-free. 
In Ref.~\cite{Cox:2017eme}, same-sign di-muon and two boosted $W_h$ jet channel has been considered.
We will, therefore, consider the same-sign dilepton channel and find out the parameter region of our model that can be probed at the HL-LHC. As a passing comment, the opposite-sign dilepton signature is important in the context of heavy pseudo-Dirac RHNs that can be present in the inverse seesaw models of neutrino mass generation.

\item \textbf{Tri-lepton:} The tri-lepton signature can arise from the following decay
\begin{eqnarray}
pp \to Z'\to N_1N_1+N_2N_2 \left\{\begin{array}{ccl}
  &\to & (W^\pm_\ell\ell^\mp)(W^\pm_h\ell^\mp) \\
   &\to & (W^\pm_h\ell^\mp)(Z_\ell\nu) 
\end{array}\right\}\ .
\end{eqnarray}
The trilepton signature is investigated in Refs.~\cite{Kang:2015uoc,Accomando:2017qcs}
where they obtained good discovery prospects of this channel over the background. There are also possibilities of four-lepton signatures which have been explored in~\cite{Huitu:2008gf} in the context of $\mathrm{U}(1)_{B-L}$ model. 

\end{itemize}

\noindent
If the light-heavy neutrino mixing angle is very small, the RHNs become long-lived which can lead to a unique displaced vertex (DV) signature which needs special analysis treatments. In~\cite{Deppisch:2019kvs,Das:2019fee,Chiang:2019ajm}, the DV signature is discussed in great detail in the RHN pair production channel through $Z'$ decay. For resonant leptogenesis to occur, the heavy RHNs are needed to be quasi-degenerate and their difference in mass should be of the same order as their decay widths. In the leptogenesis-favored parameter space, the decay widths of the RHNs are, therefore, very small and might lead to a DV signature. However, we have confirmed that our interesting regions of model parameter space are fairly safe from the DV signature. This is because since the masses of the RHNs are TeV-scale order, they are not boosted when produced and their lifetimes are not enhanced due to the time-dilation effect.

\noindent
Before we proceed further, we mention the packages we have used in our analysis. We have implemented the model Lagrangian in \texttt{FeynRules}~\cite{Alloul:2013bka} which gives model files~\cite{Degrande:2011ua} for \texttt{MadGraph}~\cite{Alwall:2014hca} event generator. We estimate the $Z'$ production cross section at the 14 TeV LHC using NNPDF2.3LO parton distribution functions~\cite{Ball:2012cx}.

\begin{figure*}[]
\captionsetup[subfigure]{labelformat=empty}
\subfloat[\quad\quad(a)]{\includegraphics[height=7cm,width=7cm]{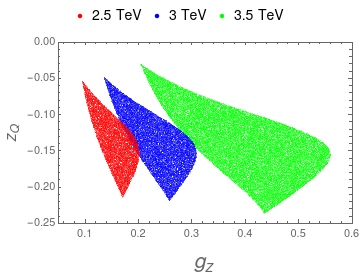}\label{fig:7a}}\quad
\subfloat[\quad(b)]{\includegraphics[height=6.5cm,width=8.5cm]{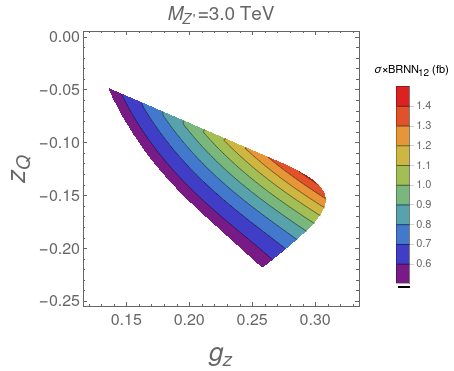}\label{fig:7b}}\\
\subfloat[\quad\quad(c)]{\includegraphics[height=7cm,width=7cm]{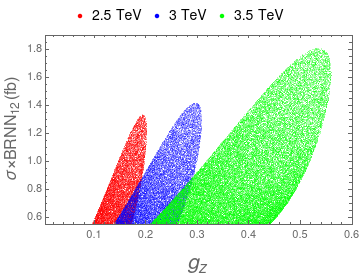}\label{fig:7c}}\quad
\subfloat[\quad\quad(d)]{\includegraphics[height=7cm,width=7cm]{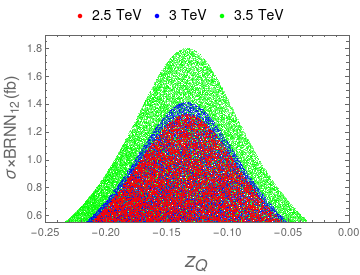}\label{fig:7d}}
\caption{In (a) and (b), we show the region in $z_Q-g_z$ plane which can be probed at the HL-LHC through the same-sign dilepton channel originates from the RHN pair production through $Z'$. In (c) and (d), the corresponding $\sg_{pp\to Z'}\times \textrm{BR}_{Z'\to N_{1,2}N_{1,2}}$ regions are shown.}
\label{fig:7}
\end{figure*}

\subsection{Constraints and prospects}
The relevant constraints that will restrict the parameter space of our model are extensively discussed in our previous work~\cite{Choudhury:2020cpm}. Here, we briefly summarize the latest data that have been used to obtain the allowed parameter space.
Although the $\sg_{pp\to Z'}\times\textrm{BR}_{Z'\to\ell\ell}$ is suppressed around $z_Q\sim -0.15$, the strongest constraints around that point come from the dilepton resonance search data at the LHC~\cite{Aad:2019fac,Sirunyan:2021khd}. Similar to dilepton, $\sg_{pp\to Z'}\times\textrm{BR}_{Z'\to jj}$ is also suppressed around 
$z_Q\sim -0.15$. The LHC dijet resonance search data~\cite{Aad:2019hjw,Sirunyan:2019vgj} can also restrict the parameter space but always give a relaxed bound compared to the bound coming from the dilepton data. Dijet bound can become important over dilepton bound if the $Z'\to jj$ BR is very large compared to $Z'\to \ell\ell$ BR (e.g. this can happen in the case for leptophobic $Z'$~\cite{Arun:2022ecj}). In our model, the $Z'\to jj$ BR is slightly bigger than $Z'\to \ell\ell$ BR for $Z_Q\gtrsim 0$ but not too big to become more important than the dilepton exclusion.
Another constraint might arise from the LEP data~\cite{Schael:2013ita} through the off-shell $Z'$ contribution to the $e^+e^-\to \bar{f}f$ processes. As the $Z'$ couplings to the leptons and quarks in our model are small, the LEP constraints do not play any major role in our analysis.
Two other low energy observables namely the $Z$-width and the $T$-parameter measurements ~\cite{Zyla:2020zbs} can play a subdominant role in obtaining bounds and are included
in our analysis.

\bigskip
\noindent
\textbf{Parameter scan:} To obtain the allowed parameter space of our model, we have used the above-mentioned data and performed a random scan over the two free parameters - 
$g_z$ and $z_Q$ for fixed benchmark values of $M_{Z'}=2.5,~3.0,~3.5$ TeV. As mentioned earlier since the dilepton resonance search data provides the most stringent bounds, we recast the 95\% confidence level upper limit on $\sigma\times BR$ obtained in~\cite{Aad:2019fac,Sirunyan:2021khd} as functions of $M_{Z'}$. After obtaining the allowed parameter space, we try to find out the parameter region which can be probed with our desired same-sign dilepton signature at the HL-LHC. Since this channel is almost background free, we naively expect that the observation of 50 signal events might be enough to achieve $5\sg$ discovery significance. We estimate the number of signal events in the following way.
\begin{equation}
\mc{N}_S =\sg_{pp\to Z'}\times BR_{Z'\to N_{1,2}N_{1,2}}\times BR^{SS}_{N_{1,2}\to
W\ell}\times (BR_{W\to jj})^2\times\varepsilon\times \mc{L}
\end{equation}
Here, $\sg_{pp\to Z'}$ is the NLO corrected production cross section of $Z'$ at the 14 TeV LHC with QCD $K$-factor is taken to be $1.3$. Both $Z'\to N_1N_1$ and $Z'\to N_2N_2$ are included in $BR_{Z'\to N_{1,2}N_{1,2}}$. The same-sign dilepton BRs of RHNs are denoted by 
$BR^{SS}_{N_{1,2}\to
W\ell}$. We have seen that in a large region of the free Yukawa couplings, BRs of RHNs to $W\mu$ mode is large while the $We$ mode is suppressed. We have only considered the hadronic decays of $W$ which has a BR of 67.6\%. To make a conservative estimate, we have also included a selection cut-efficiency of our signal including detector effects. This turned out to be around 30\% from our basic analysis. Finally, the number of events is estimated for $\mc{L}=3$ ab$^{-1}$ integrated luminosity.
In Fig.~\ref{fig:7}, we have shown our results of random scan over $g_z$ and $z_Q$ where we have chosen three benchmark masses of $Z'$. These regions can be probed at the HL-LHC in the same-sign dilepton channel originating from the RHN pair production through $Z'$ and are allowed by the dilepton resonance search data. As we increase $M_{Z'}$, to achieve the same sensitivity, we need a higher $g_z$ value to enhance the $Z'$ production cross section. Note that the limit of $g_z$ is also mass-dependent - the heavier the mass is the higher the value of $g_z$ is allowed. This is the reason why the green regions which are for $M_{Z'}=3.5$ TeV in Figs.~\ref{fig:7a}, \ref{fig:7c} and \ref{fig:7d} are bigger compared to lower benchmark masses.

\section{Summary and conclusions}\label{sec:conclusion}
We have investigated a BSM scenario augmented by a $\mathrm{U}(1)$ extension which offers neutrino mass generation through certain HDOs. In doing so, the SM fermion sector is extended by the inclusion of three RHNs, among which two RHN states essentially belong to the TeV-scale. Keeping the third RHN state in the keV scale is manifested in the motivation of identifying it as a potential dark matter candidate. The heavier RHNs are supposed to have a quasi-degenerate mass spectrum which is essentially kept to be of the order of their decay width ($\mathcal{O}$(keV)). This choice made this framework viable for explaining the BAU through the mechanism of resonant leptogenesis. Being a TeV scale model, the framework also turns out to be instrumental for studying flavored leptogenesis, which is successfully achieved.  We have also examined the possible constraints on the key parameters that bring out non-zero lepton asymmetry in this model. In this model, the low energy CP phases (the Dirac, $\delta$ and Majorana, $\alpha$  phases) are the crucial ingredients that drive the flavored lepton asymmetries towards a nonzero value. However, the aforementioned amount of degeneracy among the two heavier RHNs controls the order of lepton asymmetry and hence appears to be essential rather than a choice. We have noticed that a major contribution to the final asymmetry is provided by the $\mu$ and $\tau$ flavors having a range from $10^{-5} - 10^{-1}$. For the allowed model parameter space we have obtained the asymmetry due to the electron flavor ranging from $10^{-12} - 10^{-10}$. On the other hand, the observed baryon-to-photon ratio has put restrictions on the parameter space for the low-energy phases that enter into the PMNS matrix. A large region of the allowed Yukawa parameter space of this model can be excluded on the basis of the choice of a lepton asymmetry value of the correct sign. After we numerically obtain the magnitude of lepton asymmetry due to each flavor, we feed them in the prescription which computes the baryon to photon ratio. To get the restrictions on low energy parameter space we make use of the analytically approximated solution of the BEQs. However, to have a clear inspection of the time evolution of the generated asymmetry we have solved the flavored BEQs and computed the final asymmetry yield. It is to mention here that, the allowed values of the Yukawa parameters bring the washout scenario strictly to be strong. The strong washout regime also appears to be favored by the present evidence for neutrino masses.

In addition to accounting for the observed BAU, this model has another striking feature of getting tested at the colliders (LHC or future planned). To execute such an objective we perform the investigation through the RHN pair production. An interesting possibility of RHN pair production can occur in our model through the $pp\to Z'\to N_{1,2}N_{1,2}$ process. Since the RHNs are Majorana in nature, their same-sign dilepton decay leads to a smoking-gun signature of this channel. We have performed a prospect study of this channel and found the region of parameter space that can be probed at the HL-LHC. The prospect of this channel is usually low in a wide class of anomaly-free $\mathrm{U}(1)$ extensions of the SM. For example, in the commonly considered $\mathrm{U}(1)_{B-L}$ model, the $Z'$ BR to RHNs are small and as a result the dilepton resonance search data severely restrict the parameter space. This leaves very little or no parameter space left for observing the pair production of RHNs through $Z'$ that can lead to the desired same-sign dilepton signature. In our model, at around $z_Q\sim -1/3$, $Z'\to\ell\ell$ BR is suppressed and $Z'\to N_{1,2}N_{1,2}$ BR is as high as 30\%. This helps us bypass the dilepton constraints and leaves a substantial region of parameter space open for observing the signature of our interest.

In a model with HDOs, where the smallness of the neutrino mass is attributed to the suppressive power of the HDOs, the same HDOs do arrange sufficient leptonic asymmetry with the same order of small to moderate Yukawa coupling values. The common feature in both of these is the presence of RHNs; the heavier quasi-degenerate duo among them do the job. These same RHNs on the collider side, facilitate the emergence of a novel same-sign dilepton channel. This connection triad is unique to the use of very specific HDOs, guided by anomaly free nature of the gauge structure.

\section*{Acknowledgment}
The authors gratefully acknowledge Prof. Debajyoti Choudhury for having a useful discussion. A.M. acknowledges the financial support provided by SERB-DST, Govt. of India through the project EMR/2017/001434. A.M. also acknowledges the post-doctoral fellowship provided by the Saha Inst. of Nuclear Physics, Kolkata. A.M. wants to acknowledge Abhijit Kumar Saha for numerous discussions in various stages of solving the BEQs in MATHEMATICA. T.M. is supported by the SERB, India under research grant CRG/2018/004889 and an intramural grant from IISER-TVM.  S.S. acknowledges research facilities provided by Vivekananda Centre for Research (recognized under the University of Calcutta). K.D. acknowledges Council for Scientific and Industrial Research (CSIR), India for JRF/SRF fellowship with the award letter no. 09/045(1654)
/2019-EMR-I.

\bigskip
\appendix

\section{Quasi-degenerate spectrum and discrete symmetry}
\label{app:flavsym}
Here we briefly discuss some of the possible ways one can adopt in order to address the required degeneracy among the RHN mass states which is demanded by the success of resonant leptogenesis.  
\subsection{Resonant Leptogenesis with $S_3$ flavor Symmetry}
The effective HDO theory will be having a quasi-degenerate spectrum for the RHN masses viable for the resonant leptogenesis scenario, with the imposition of the $S_3$ flavor symmetry (see e.g. \cite{Ishimori:2010au} for the details of $S_3$ symmetry). Implications of such flavor symmetries in order to address the degeneracy among the RHN mass states can also be found in \cite{Borah:2017qdu,Mishra:2019sye}.
Anomaly-free Frogatt Nielsen charge assignment is arranged to write the Lagrangian above, where all three RHNs are SM singlets with $\mathrm{U}(1)_{\rm FN}$ charges $ 4, 4, -5$. We are not going to alter the Frogatt Nielsen charge assignment of the fields of theory. 

Now, two heavier RHNs $N_1, N_2$ can transform like a doublet ($\bf 2$) under the $S_3$ group while the ultra-light third neutrino $N_3$ is an $S_3$ singlet ($\bf 1$). All other fields introduced earlier remain to be $S_3$ singlets. Purposefully we introduce one $S_3$ scalar doublet $(\chi_3 , \chi_4)^T$. 

%Under the $S_3$ algebra the two dimensional IRs decompose as 
The tensor product of 2-dimensional Irreducible representations of the $S_3$ can be cast into the following form (see \cite{Chen:2004rr,Ma:2015raa}).   
$$ {\bf 2} \otimes {\bf 2} = {\bf 1} \oplus {\bf 1^{\prime}} \oplus {\bf 2}.$$ 
Here, $(N_1 \ \ N_2)^T$ as well as $(\bar{N}_1^c \ \ \bar{N}_2^c)^T$ form the $\bf 2$'s of $S_3$. 
The singlet part ($\bf 1$) will be 
$$ (\bar{N}_1^c N_1 + \bar{N}_2^c N_2 ).$$ So the mass terms of the two heavier right-handed fermions will appear with a single mass term as
$$ y \frac{S_2^2}{\Lambda} (\bar{N}_1^c N_1 + \bar{N}_2^c N_2 ).$$ This is in  clear contrast with the same HDO setup without the $S_3$ symmetry where the  $\bar{N}_1^c N_1, \bar{N}_2^c N_2$ terms appear with different Yukawa couplings. The ${\bf 1^{\prime}}$ of the $S_3$ group is $ (\bar{N}_1^c N_2 + \bar{N}_2^c N_1 )$ which is not allowed in the model and so the off-diagonal terms in the $N_1-N_2$ sector which could have been in same orders of magnitude as the diagonal ones are not present. That makes the $N_1 - N_2$ sector exactly diagonal. Exactly diagonal RHN sector with a minuscule breaking is the basic building block of lifting the exact degeneracy among the RHN masses. Now for the degeneracy breaking between the $N_1$ and $N_2$ we have to introduce a new set of scalars apart from the already introduced $S_1, S_2$. We introduce two scalars $\chi_1, \chi_2$ that form a ${\bf 2}$ 
under $S_3$ with gauge charges $$ z_{\chi_1} = z_{\chi_2} = - \frac{3}{4},$$ while these scalars are singlets under the SM symmetry. With this set up the $N_{1/2}-N_3$ sector interaction terms will look like (keeping only the ${\bf 1}$ under $S_3$) 
$$ y_{3} \ (N_1 \chi_1 + N_2 \chi_2) N_3 \frac{S_1^3 S_2^{\*}}{\Lambda^4}. $$ The $N_3$ mass terms remain the same as it is a singlet under $S_3$. 
\begin{table*}
\begin{center}
\begin{tabular}{| c | c | c | c|c|  }
  \hline
BP  &$\epsilon_e $ & $\epsilon_\mu$ & $\epsilon_\tau$&  $ \eta_B$  \\
\hline 
I &   $ -1.4 \times 10^{-5}$  & $-7.46\times 10^{-6} $   & $-2.1 \times \times 10^{-5}$& $3.2\times 10^{-10}$ \\
  \hline

II &   $ -1.06 \times 10^{-5} $  & $-1.1\times 10^{-5} $   & $-1.36 \times 10^{-5}$& $3.18 \times 10^{-10}$ \\
\hline
III &   $ - 1.38\times 10^{-5} $  & $- 6.13\times 10^{-6} $   & $-1.68 \times 10^{-5}$& $3.21 \times 10^{-10}$ \\
\hline
\end{tabular}
\caption{Benchmark values for the lepton asymmetry associated with each lepton flavor and resulting baryon asymmetry.}
\label{tab:etabS3}
\end{center}
\end{table*}

The Yukawa terms involving $N_3$ and other HDOs involving only the SM fields remain as stated in the main text. The Yukawa couplings involving the 
left-handed leptons $L_i$ and RHNs $N_1, N_2$ get modified through 
$$ y_{\ell} \bar{L}_{\ell} (N_1 \chi_1 + N_2 \chi_2) \tilde{H} \frac{S_1^3}{\Lambda^4}.$$
With the above terms in the Dirac and heavy Majorana mass matrices a resonant enhancement in the lepton asymmetry can be achieved naturally. In the above (in Table~\ref{tab:etabS3}) we have provided some representative values of the lepton and baryon asymmetry yield that we obtain under this $S_3$ realization. These benchmark points (BP) are chosen from the scan such that Yukawa couplings are moderate and we check the asymmetries for those values.

In a similar way, one can also consider an $U(2)$ flavor symmetry in the RHN sector. As before, $(N_1 \ \ N_2)^T$, $(\bar{N}_1^c \ \ \bar{N}_2^c)^T$ form doublet under $U(2)$ whereas $N_3$ and all other fields are singlets of $U(2)$. Cross terms like $N_1N_2$ type in Eq.~\eqref{nu_mass_our} will be removed after this imposition. To keep other terms involving RHNs $U(2)$ invariants, we can follow the same prescription as we have done above for $S_3$ by introducing a scalar doublet $(\chi_3 , \chi_4)^T$ with $U(2)$ charge $-3/4$. However, we need to break this $U(2)$ by the scalar vevs in order to obtain similar interactions in Eq.~\eqref{nu_mass_simpl}.

\subsection{Resonant Leptogenesis with $N_1-N_2$ exchange symmetry}
First we consider the transformation $N_1 \rightarrow N_2$ and $N_2 \rightarrow -N_1$. So $\bar{N_1^c}N_1 \rightarrow \bar{N_2^c}N_2$ and $\bar{N_2^c}N_2 \rightarrow \bar{N_1^c}N_1$, whereas $\bar{N_1^c}N_2 + \bar{N_2^c}N_1$ goes to negative of itself. Thus we are left with only the diagonal terms. The Dirac terms involve only the scalar $S_1$. We introduce another scalar $\tilde{S}_1$ with the same $\mathrm{U}(1)_z$ charge of $-3/4$ as $S_1$ such that they transform as $S_1 \rightarrow c_1 \tilde{S}_1$ and $\tilde{S}_1 \rightarrow c_2 S_1$. Therefore the Dirac term $\bar L_{L} N_{R} \tilde{H} S_1^{4}$ transforms to $\bar L_{L} N_{R} \tilde{H} S_1^{a} \tilde{S}_1^{b}$ such that $(a+b)=4$. Therefore under the symmetry transformation of the RHNs and the scalars, we have:
\begin{equation}\nonumber
\begin{split}
\bar L_{L} N_1 \tilde{H} S_1^{a} \tilde{S}_1^{b} \rightarrow \bar L_{L} N_2 \tilde{H} S_1^{b} \tilde{S}_1^{a} c_1^a c_2^b~, ~~~(a+b)=4 \\
\bar L_{L} N_2 \tilde{H} S_1^{d} \tilde{S}_1^{e} \rightarrow -\bar L_{L} N_1 \tilde{H} S_1^{e} \tilde{S}_1^{d} c_1^d c_2^e~, ~~~(d+e)=4 
\end{split}
\end{equation}
Comparing with the original terms we have:
\begin{equation}
\nonumber
a=e~, ~~~b=d~, ~~~c_1^a c_2^b =1~, ~~~c_1^d c_2^e =-1 
\end{equation}
This implies $c_1^4c_2^4=-1$, resulting in the choice $a=4$, $b=0$, $d=0$ and $e=4$ along with $c_1^2=1$ and $c_2^2=i$. This choice makes sure that the Dirac terms involving $N_1$ and $N_2$ go to each other after transformation under this discrete symmetry. This degeneracy can be softly broken by allowing for terms higher in dimension. With these choices of parameters, $(a,b,c,d,c_1,c_2)$ the other mass terms follow trivially.
%%%%%%%%%%%%%%%%%%%%%%%%%%%%%%%%%%%%%%%%%%%%%%%%%%%%%%%%%%%%%%%

\section{Scattering and decay rates of RHNs}
\label{sec:Appendix B}
The following expressions for the RHN decay and scattering rates have been taken from Refs.~\cite{Pilaftsis:2005rv, Dolan:2018qpy}.
The BEQs \ref{beq1} and \ref{beq2} described in section \ref{sec:beq} require the following definition of some of the variables as mentioned below. 
\begin{equation}\
\label{rescaled}
z = \frac{m_{N_1}}{T}\,, \;\;x = \frac{s}{m_{N_1}^2}\,, \;\; 
a_i = \left(\frac{m_{N_i}}{m_{N_1}}\right)^2\,, \;\; 
a_r = \left(\frac{m_{\rm IR}}{m_{N_1}}\right)^2 \simeq 10^{-5}\,, \;\;
c_i\ =\ \left(\,\frac{\Gamma_{N_i}}{m_{N_1}}\,\right)^2\; .
\end{equation}
with  $s$ being the Mandelstam variable, and the infrared mass regulator for the t-channel $m_{\rm{IR}} = m_{\Phi} / M_{N_1}$,
whose value is set to $10^{-5}$ as mentioned in \cite{Luty:1992un}.

The total decay width $\Gamma_{N_i}$ of the RHNs is given by
\begin{equation}
\Gamma_{N_i}\ =\ \sum_{l=1}^{3} \Gamma^{\ell}_{N_i} =\ \frac{m_{N_i}}{8\pi}\
\sum_{l=1}^{3} y_{i\ell}^{*} y_{i\ell} .
\end{equation}

The collision terms for $1\to 2$ and $2\to 2$ processes which appear in Eqs.~\eqref{beq1} and \eqref{beq2} are calculated in Ref.~\cite{Pilaftsis:2005rv}.
The rate for a generic process $X\to   Y$ and its conjugate counterpart $\overline{X}\to \overline{Y}$ is defined as $\gamma^X_Y$.
For   the  $1\to   2$   process,\   $N_i\to   L\Phi$  or   $N_i\to
L^C\Phi^\dagger$, 
$\gamma^{X}_{Y}$ is given by
\begin{eqnarray}
\gamma^{N_i}_{L_\ell \Phi}\ = \frac{m_{N_1} m_{N_i}^{2} \Gamma_{N_{i}}^{\ell} }{\pi^2\, z}\ K_1(z \sqrt{a_i})\,,
\end{eqnarray}
in terms of the rescaled variables of Eq.~\eqref{rescaled} where$K_n(z)$ is an
$n$th-order modified Bessel function.
The $2 \to 2$ processes can be divided into $\Delta L = 0,1,2$ cases, each contributing to the washout of lepton flavor at different rates.
For a generic $2 \to 2$ process the collision term is calculated through
\begin{equation}
  \label{2t2g}
\gamma^{X}_{Y} = \frac{m^4_{N_1}}{64\,\pi^4 z} \int\limits_{x_{\rm thr} }^\infty\! dx \sqrt{x} \; K_1(z\sqrt{x}) \; \sigma^{X}_{Y}(x)\ ,
\end{equation}
where $x_{\rm thr}$ corresponds to the threshold, $\rm{Min}(m(X),m(Y))$. These processes can be computed by substituting each reduced cross section $\sigma^{X}_{Y}$, numerically interpolating \ref{2t2g} over a range of z values and including in the numerical BEQs.

Following are the cross sections for the possible $\Delta L = 1$ processes,
\begin{eqnarray}
  \label{del1-1}
\sigma^{N_i L_\ell}_{Qu^C} & = & \frac{3 y_{t}^{2}}{4\pi}   \left( y_{i \ell}^{*} y_{i \ell} \right)\left(\frac{x-a_i}{x}\right)^2\;,\nonumber\\
  \label{del1-2}
\sigma^{N_iu^C}_{L_\ell Q^C} & = & \sigma^{N_iQ}_{L_\ell u}\nonumber\\ 
&=& \frac{3 y_{t}^{2}}{4\pi}  \left( y_{i \ell}^{*} y_{i \ell} \right)\left(1-\frac{a_i}{x}+\frac{a_i}{x}
\ln\left(\frac{x-a_i+a_r}{a_r}\right)\right),\nonumber\\
  \label{del1-3}
\sigma^{N_iV_\mu}_{L_\ell \Phi} \!&=&\! 
\frac{3 g^2}{8\pi\,x} \,
\left(y_{i \ell}^{*} y_{i \ell} \right)
\left(\frac{(x+a_i)^2}{x-a_i + 2 a_r}\,
\ln\left(\frac{x-a_i + a_r}{a_r}\right)\right),\nonumber\\
\label{del1-4}
\sigma^{N_i L_\ell}_{\Phi^\dagger V_\mu} \!&=&\!
\frac{3 g^2}{16\pi\, x^2}\left(y_{i \ell}^{*} y_{i \ell}  \right)\left(5x-a_i)\,(a_i -x) +
2(x^2+xa_i - a^2_i)\,
\ln\left(\frac{x-a_i + a_r}{a_r}\right)\right),\nonumber\\
\label{del1-5}
\sigma^{N_i \Phi^\dagger}_{L_\ell V_\mu} \!&=&\!
\frac{3 g^2}{16\pi\, x^2}\left(y_{i \ell}^{*} y_{i \ell}  \right) (x-a_i) \left( x-3a_i + 4a_i\,
\ln\left(\frac{x-a_i+a_r}{a_r}\right)\right)\,.
\end{eqnarray}
%For the scatterings involving gauge bosons, only the $SU(2)_{L}$ processes were considered as the $U(1)_{Y}$ processes are expected to be subdominant.

The necessary $\Delta L = 2$ processes and cross-sections are,
\begin{eqnarray}
\label{del2-1}
\sigma^{ L_\ell \Phi}_{L_k^C\Phi^\dagger} = 2 \sum_{i,j=1}^{2} \textrm{Re}\Big\lbrack\left(y_{j \ell} y_{j k} y_{i \ell}^{*} y_{i \ell}^{*}\right) \mathcal{A}^{ss}_{ij} & + &  \left(y_{j \ell} y_{j k} y_{i \ell}^{*} y_{i k}^{*}\right) \mathcal{A}^{tt}_{ij}\nonumber\\
& + & \left(y_{i \ell}^{*} y_{i k} y_{j \ell}^{*} y_{j k} + y_{i \ell} y_{i k}^{*} y_{j \ell} y_{j k}^{*}\right)\mathcal{A}^{(st)*}_{ij}\Big\rbrack\nonumber\\
\label{del2-2}
\sigma^{L_{\ell} L_{k} }_{\Phi^\dagger\Phi^\dagger} \ = \
\sum_{i,j=1}^{2} {\rm Re} \left(y_{j \ell} y_{j k} y_{i \ell}^{*} y_{i k}^{*} \right) \mathcal{B}_{ij}\;,
\end{eqnarray}
where
\begin{eqnarray}
\label{aij}
\mathcal{A}^{(ss)}_{ij} \!\!&=&\!\! \left\{
\begin{array}{cc}
 0 &\quad (i=j)\,,\\
\frac{\displaystyle x\sqrt{a_i\,a_j}}{\displaystyle 
4\pi P^*_i P_j}\ &\quad (i\neq j)\,,
\end{array} \right. \nonumber\\
\mbox{}\nonumber\\
\mathcal{A}^{(tt)}_{ij} \!\!&=&\!\! \left\{
\begin{array}{cc}
\frac{a_i}{2\pi x}\Bigg[\, \frac{x}{a_i}\ -\
\ln\bigg(\frac{x+a_i}{a_i}\bigg)\, \Bigg], &\quad (i=j)\,,\nonumber\\
\frac{\sqrt{a_i\,a_j}}{2\pi x\:(a_i-a_j)}\left[\,
(x+a_j)\,\ln\bigg(\frac{x+a_j}{a_j}\bigg)\ -\
(x+a_i)\,\ln\bigg(\frac{x+a_i}{a_i}\bigg)\, \right], &\quad (i\neq j)\,,
\end{array} \right. \nonumber\\
\mbox{}\nonumber\\
\mathcal{A}^{(st)}_{ij} \!\!&=&\!\! 
\frac{\sqrt{a_i\,a_j}}{2\pi P_i}\left[\, 1\ -\ \frac{x+a_j}{x}\,
\ln\bigg(\frac{x+a_j}{a_j}\,\bigg)\right]\,,\nonumber\\
\mbox{}\nonumber\\
\mathcal{B}_{ij} \!\!&=&\!\! \left\{
\begin{array}{cc}
 \frac{1}{2\pi}\left[\,\frac{x}{x+a_i}\: +\: \frac{2\,a_i}{x+2a_i}
\ln\left(\frac{x+a_i}{a_i}\right)\,\right], &\quad (i=j)\,,\nonumber\\
\frac{\sqrt{a_i\,a_j}}{2\pi}
\left[\, \frac{1}{a_i-a_j}
\ln\left(\frac{a_i(x+a_j)}{a_j(x+a_i)}\right)\: +\:
\frac{1}{x+a_i+a_j}
\ln\left(\frac{(x+a_i)(x+a_j)}{a_i\,a_j}\right)\,\right]. &\quad (i\neq j)\,,
\end{array} \right.  \nonumber\\
\end{eqnarray}
and
\begin{equation}
  \label{Pi}
P^{-1}_i (x) \ =\ \frac{1}{x-a_i+i\sqrt{a_i c_i}}\ .
\end{equation}

Following are the cross sections for possible  $\Delta L = 0$ processes, taken from \cite{Pilaftsis:2005rv}.
\begin{eqnarray}
\sigma^{L_\ell\Phi}_{L_k \Phi} \!& = &\! 
\sum_{i,j=1}^2 \left(y_{i k}^* y_{i \ell} y_{j k} hy_{j \ell}^* + 
y_{i k } y_{i \ell}^* y_{j k}^* y_{j \ell}\right)\mathcal{C}_{i j}\nonumber\\
\sigma^{L_\ell\Phi^\dagger}_{L_k \Phi^\dagger} \!& = &\!
\sum_{i,j=1}^2 \mathrm{Re} \Big( y_{p\ell}^*\,y_{q\ell}\, 
y_{p k}\,y_{q k}^* \Big)\,\mathcal{D}_{i j}\,,\nonumber\\
\sigma^{L_\ell L^C_k}_{\Phi^\dagger \Phi} \!& = &\!
\sum_{i,j=1}^2 \mathrm{Re} 
\Big( y_{p \ell}^*\,y_{ q \ell}\, 
y_{p k }\,y_{q k}^* \Big)\,\mathcal{E}_{i j}\,,
\end{eqnarray}
where
\begin{eqnarray}
\label{Cab}
\mathcal{C}_{i j} \!\!&=&\!\! \left\{
\begin{array}{lc}
0         &\quad (i = j)\\
\frac{\displaystyle x\sqrt{a_i a_j}}{\displaystyle 
4\pi P^*_i P_j}\ &\quad (i\neq j)
\end{array} \right.\nonumber\\
\mathcal{D}_{i j} \!\!&=&\!\! \left\{
\begin{array}{lc}
\frac{a_i}{\pi x}\:
\Bigg[\,\frac{x}{a_i} - \ln\Bigg(\frac{x+a_i}{a_i}\Bigg)         &\quad (i = j)\\
\frac{\sqrt{a_i a_j}}{\pi
x (a_i - a_j)}\: 
\Bigg[\, (x+a_j)\ln\Bigg(\frac{x+a_j}{a_j}\Bigg)-(x+a_i)
\ln\Bigg(\frac{x+a_i}{a_i}\Bigg)\, \Bigg] &\quad (i\neq j)
\end{array} \right.\nonumber\\
\mathcal{E}_{i j} \!\!&=&\!\! \left\{
\begin{array}{lc}
\frac{x}{\pi (x+a_i)}         &\quad (i = j)\\
\frac{\sqrt{a_i a_j}}{\pi
(a_i - a_j)}\:
\ln\Bigg(\frac{a_i (x+a_j)}{a_j (x+a_i)}\Bigg) &\quad (i \neq j)
\end{array} \right.
\end{eqnarray}

The scattering rates used in Eqs.~\eqref{beq1} and \eqref{beq2} are then calculated as functions of the collision terms above,
\begin{eqnarray}
\Gamma^{D(i \ell)} & = & \frac{1}{n_\gamma}\
\gamma^{N_{i}}_{L_\ell\Phi}\nonumber\\
\widehat{\Gamma}^{D(i \ell)} & = &
\frac{1}{n_\gamma}\left(1+\frac{4}{21}\,\frac{\eta_{\Delta L}}{\eta_{\Delta L_\ell}} \right)\,
\gamma^{N_{i}}_{L_\ell\Phi}\nonumber\\
\widetilde{\Gamma}^{D(i \ell)} & = &
\frac{1}{n_\gamma}\left(1+\frac{4}{21}\,\frac{\eta_{\Delta L}}{\eta_{\Delta L_\ell}} \right)\,
\gamma^{N_{i}}_{L_\ell\Phi}\nonumber\\
\Gamma^{S (i \ell)}_{\rm Y} & = & \frac{1}{n_\gamma}\
\left(\, \gamma^{N_{i} L_\ell}_{Q u^C} +  \gamma^{N_{i}
u^C}_{L_\ell Q^C} + \gamma^{N_{i} Q}_{L_\ell u}\, \right),\nonumber\\
\widehat{\Gamma}^{S(i \ell)}_{\rm Y} &=& \frac{1}{n_\gamma}\
\biggl[\left(-\frac{\eta_{N_{i}}}{\eta^{\rm eq}_{N_{i}}}
+\frac{4}{21}\,\frac{\eta_{\Delta L}}{\eta_{\Delta L_\ell}}\right)
\gamma^{N_{i} L}_{Q u^C} +
\left(1+\frac{1}{9}\,\frac{\eta_{\Delta L}}{\eta_{\Delta L_\ell}}
-\frac{5}{63}\,\frac{\eta_{N_{i}}}{\eta^{\rm eq}_{N_{i}}}
\frac{\eta_{\Delta L}}{\eta_{\Delta L_\ell}}\right)
\gamma^{N_{\ell} u^C}_{L Q^C}\nonumber\\
& & +\left(1+\frac{5}{63}\,\frac{\eta_{\Delta L}}{\eta_{\Delta L_\ell}}
-\frac{1}{9}\,\frac{\eta_{N_{i}}}{\eta^{\rm eq}_{N_{i}}}
\frac{\eta_{\Delta L}}{\eta_{\Delta L_\ell}}\right)
\gamma^{N_{i} Q}_{L u}\biggr]\nonumber\\
\widetilde{\Gamma}^{S(i \ell)}_{\rm Y} &=& \frac{1}{n_\gamma}\
\biggl[\left(\frac{\eta_{N_{i}}}{\eta^{\rm eq}_{N_{i}}}
+\frac{4}{21}\,\frac{\eta_{\Delta L}}{\eta_{\Delta L_\ell}}\right)
\gamma^{N_{i} L}_{Q u^C} +
\left(1+\frac{1}{9}\,\frac{\eta_{\Delta L}}{\eta_{\Delta L_\ell}}
+\frac{5}{63}\,\frac{\eta_{N_{i}}}{\eta^{\rm eq}_{N_{i}}}
\frac{\eta_{\Delta L}}{\eta_{\Delta L_\ell}}\right)
\gamma^{N_{i} u^C}_{L Q^C}\nonumber\\
& & + \left(1+\frac{5}{63}\,\frac{\eta_{\Delta L}}{\eta_{\Delta L_\ell}}
+\frac{1}{9}\,\frac{\eta_{N_{i}}}{\eta^{\rm eq}_{N_{i}}}
\frac{\eta_{\Delta L}}{\eta_{\Delta L_\ell}}\right)
\gamma^{N_{i} Q}_{L u}\biggr]\nonumber\\
\Gamma^{S(i \ell)}_{\rm G} & = & \frac{1}{n_\gamma}
\biggl(\gamma^{N_{i} L_\ell}_{\Phi^\dagger V_\mu} +
\gamma^{N_{i} V_\mu}_{L_\ell\Phi} +
\gamma^{N_{i}\Phi^\dagger }_{L_\ell V_\mu}\biggr)\nonumber\\
\widehat{\Gamma}^{S(i \ell)}_{\rm G} &=& \frac{1}{n_\gamma}\ 
\biggl[\left(-\frac{\eta_{N_{i}}}{\eta^{\rm eq}_{N_{i}}}
+\frac{4}{21}\frac{\eta_{\Delta L}}{\eta_{\Delta L_\ell}}\right)
\gamma^{N_{i} L}_{\Phi^\dagger V_\mu}
+\left(1+\frac{4}{21}\,\frac{\eta_{\Delta L}}{\eta_{\Delta L_\ell}}\right)
\gamma^{N_{i} V_\mu}_{L \Phi}\nonumber\\
& & + \left(1-\frac{4}{21}\,\frac{\eta_{N_{i}}}{\eta^{\rm
eq}_{N_{i}}}
\frac{\eta_{\Delta L}}{\eta_{\Delta L_\ell}}\right)
\gamma^{N_{i}\Phi^\dagger }_{L V_\mu}\biggr]\nonumber\\
\widetilde{\Gamma}^{S(i \ell)}_{\rm G} &=& \frac{1}{n_\gamma}
\biggl[ \left( \frac{\eta_{N_{i}}}{\eta^{\rm eq}_{N_{i}}}+\frac{4}{21}\,\frac{\eta_{\Delta L}}{\eta_{\Delta L_\ell}} \right) 
\gamma^{N_{i} L}_{\Phi^\dagger V_\mu}
 +\left(1+\frac{4}{21}\,\frac{\eta_{\Delta L}}{\eta_{\Delta L_\ell}}\right)
\gamma^{N_{i} V_\mu}_{L \Phi}\nonumber\\
& & +\left(1+\frac{4}{21}\frac{\eta_{N_{i}}}{\eta^{\rm eq}_{N_{i}}}
\frac{\eta_{\Delta L}}{\eta_{\Delta L_\ell}}\right)
\gamma^{N_{i}\Phi^\dagger }_{L V_\mu}\Biggr]\nonumber\\
\Gamma^{W(i \ell)}_{\rm Y} & = & \frac{1}{n_\gamma}
\biggl[\left(2+\frac{4}{21}\frac{\eta_{\Delta L}}{\eta_{\Delta L_\ell}}\right)
\gamma^{N_{i} L}_{Q u^C} +
\left(1+\frac{17}{63}\frac{\eta_{\Delta L}}{\eta_{\Delta L_\ell}}\right)
\gamma^{N_{i} u^C}_{L Q^C}\nonumber\\
& & +\left(1+\frac{19}{63}\frac{\eta_{\Delta L}}{\eta_{\Delta L_\ell}}\right)
\gamma^{N_{\ell} Q}_{L u}\biggr]\nonumber\\
\Gamma^{W(i \ell)}_{\rm G} & = & \frac{1}{n_\gamma}\ 
\biggl[\left(2+\frac{4}{21}\,\frac{\eta_{\Delta L}}{\eta_{\Delta L_\ell}}\right)
\gamma^{N_{i} L}_{\Phi^\dagger V_\mu} +
\left(1+\frac{4}{21}\,\frac{\eta_{\Delta L}}{\eta_{\Delta L_\ell}}\right)
\gamma^{N_{i} V_\mu}_{L \Phi}\nonumber\\
& & + \left(1+\frac{8}{21}\frac{\eta_{\Delta L}}{\eta_{\Delta L_\ell}}\right)
\gamma^{N_{i} \Phi^\dagger }_{L V_\mu}\biggr]\nonumber\\
\end{eqnarray}
%%%%%%%%%%%%%%%%%%
\begin{eqnarray}
\Gamma^{\Delta L =2(\ell k)}_{\rm Y} &=& \frac{1}{n_\gamma}\ 
\biggl[\left(1+\frac{4}{21}\frac{\eta_{\Delta L}}{\eta_{\Delta L_\ell}}\right)
\left(\gamma^{L_\ell\Phi}_{\,L_k^C\Phi^\dagger} + 
\gamma^{L_\ell L_k}_{\Phi^\dagger\Phi^\dagger}\right)\biggr]\nonumber\\
\label{scatterings}
\Gamma^{\Delta L =0(\ell k)}_{\rm Y} &=& \frac{1}{n_\gamma}\ 
\biggl[\left(1+\frac{4}{21}\,\frac{\eta_{\Delta L}}{\eta_{\Delta L_\ell}}\right)
\gamma^{L_\ell\Phi}_{\,L_k\Phi}  + 
\gamma^{L_\ell\Phi^\dagger}_{L_k Phi^\dagger} + 
\gamma^{L_\ell L^C_k}_{\Phi \Phi^\dagger}\biggr].
\end{eqnarray}

\bibliographystyle{JHEP}
\bibliography{leptogenesis}
\end{document}